\documentclass[epj]{svjour}
\usepackage{latexsym}
\usepackage{array}
\usepackage{amsmath}
\usepackage{epsfig}
\usepackage{graphics}
\usepackage{booktabs}
\usepackage{feynmp}
\include{physjour.bst}
\newcommand{\tn}[1]{\ensuremath{10^{#1}}}
\newcommand{\ttn}[1]{\ensuremath{\times 10^{#1}}}
\newcommand{\neut}{\ensuremath{\ti{\chi}^0}}
\newcommand{\charg}{\ensuremath{\ti{\chi}^+}}
\newcommand{\GeV}{\mbox{\!\ GeV}}
\newcommand{\TeV}{\mbox{\!\ TeV}}

\newcommand{\fb}{\mbox{\!\ fb}}
\newcommand{\yr}{\mbox{\!\ yr}}

\newcommand{\mb}{\mbox{\!\ mb}}
\newcommand{\scm}{\mbox{\!\ cm$^{-2}$s$^{-1}$}}

\newcommand{\LV}{\mbox{$L$\hspace*{-0.5 em}/\hspace*{0.18 em}}}
\newcommand{\BV}{\mbox{$B$\hspace*{-0.65 em}/\hspace*{0.3 em}}}
\newcommand{\RV}{\mbox{$R$\hspace*{-0.6 em}/\hspace*{0.3 em}}}
\newcommand{\ET}{\ensuremath{E_T\hspace*{-1.15em}/\hspace*{0.65em}}}

\newcommand{\Lum}{\ensuremath{\mathcal{L}}}

\newcommand{\ti}[1]{\ensuremath{\tilde{#1}}}
\newcommand{\pythia}{\textsc{Pythia}}
\newcommand{\susygen}{\textsc{SusyGen}}

\newcommand{\atlfast}{\textsc{AtlFast}}
\newcommand{\herwig}{\textsc{Herwig}}

\newcommand{\isasusy}{\textsc{Isasusy}}
\newcommand{\isawig}{\textsc{Isawig}}

\newlength{\tfcapsep}
\begin{document}
\title{Searching for L-Violating Supersymmetry at the LHC}
\author{P. Z. Skands
\thanks{\emph{Now at Department of Theoretical Physics, Lund
    University, 
    S\"{o}lvegatan 14A, S-22362 Lund (zeiler@thep.lu.se)}}}
\institute{Niels Bohr Institute, Blegdamsvej 17, 2100 Copenhagen \O, Denmark}
\date{LU TP 01-32}
\abstract{
The possibility to simulate lepton  number violating supersymmetric models
has been
introduced into the recently updated \pythia\ event generator, now containing 
1278 decay channels of SUSY particles into SM particles via lepton number
violating interactions. This generator has been used in combination with the
\atlfast\ detector simulation to study the impact of
lepton number violation (\LV) on event topologies in the ATLAS detector, and
trigger menus designed for \LV-SUSY are proposed based on very general
considerations. In addition, a rather preliminary analysis is presented on 
the possibility for ATLAS to observe a signal 
above the background in several mSUGRA
scenarios, using a combination of primitive cuts and
neural networks  
to optimize the discriminating power between signal and
background events over regions of parameter space rather than at individual
points. 
It is found that a $5\sigma$ discovery is possible roughly for
$m_{1/2}<1\TeV$ and $m_0<2\TeV$ with an integrated luminosity of 30\fb$^{-1}$, 
corresponding to one year of data taking with the LHC running 
at ``mid-luminosity'', $\Lum=3\ttn{33}\scm$.
}
\maketitle
\section{Introduction}
Among the primary physics goals of the Large Hadron Collider at CERN, scheduled to turn on some time during 2006, are the exploration of the SM Higgs and top quark sectors, 
and the direct exploration of the \TeV\ scale with the emphasis on
Supersymmetry. 

The motivations for believing that Supersymmetry is
indeed a property of Nature are many, most importantly 
the natural and exhaustive 
extension of the Poincar\'{e} group 
furnished by the Supersymmetric operators \cite{haag75}, 
the possibility for exact unification of the SM gauge couplings as required
by GUTs, the requirement of Supersymmetry for anomaly-cancellation 
in string theories, and perhaps
most importantly, its providing a natural and elegant 
solution to the problem of scale hierarchy in the SM. 

For this latter problem to be solved without unnatural finetuning order by
order in perturbation theory, the sparticle masses must lie at or below the
\TeV\ scale, and so there is ample reason to believe that SUSY should be
observable at the LHC (see e.g.\ \cite{ellisbm}).  

The potential for hadron colliders to observe Supersymmetry 
has been studied in great detail (see e.g.\ \cite{atlastdr,abel00}), yet
studies tend to concentrate on the simple MSSM framework where both baryon and
lepton number conservation are imposed. In section \ref{sec:LV} we summarize
the theoretical situation,
pointing out that it is by no means obvious that
baryon and lepton number should be conserved in supersymmetric theories. 

The possibility to study lepton and baryon number violating Supersymmetry has
already been included
\cite{richardson00} in the \herwig\ event generator \cite{herwig6} and in
\susygen\ \cite{ghodbane99}, and several
studies exist for run II of the Tevatron (for reviews, see
\cite{ambrosanio98,allanach99}, latest experimental results \cite{dzerorpv})
and to a lesser extent for the LHC (see e.g.\ 
\cite{dimopoulos90,dreiner00-2,abdullin99,xi01}).  

However, it has not before been possible to study the full range of decays in
these supersymmetric models
using the \pythia\ framework (and thus the string fragmentation model), and
no dedicated LHC studies have been performed regarding triggers and
sensitivity for these theories beyond cases where the LSP was simply forced
to decay. In this paper, we go to the general case where all sparticles
(excepting the gluino) are
allowed to decay via lepton number violating couplings, introducing 1278
lepton number violating
decay channels of sparticles to particles in the \pythia\ event generator. 

For the purpose of studying the experimental signatures arising from the
lepton number violating decay channels, and to quantify to some extent the
``discovery potential'' of the LHC, some selection of benchmark points is
needed. 5 mSUGRA points and, for each of these, 
9 scenarios for the 36 \LV\ coupling strengths are presented and discussed in
section \ref{sec:mSUGRA}. 

From this point on, the analysis relies on
technical aspects of the detector design (or rather, their representation in
crude simulation algorithms), and so we are forced to distinguish between 
the ATLAS and CMS experiments. Assuming to a first approximation that these
experiments will have similar 
capabilities, we choose to concentrate on ATLAS in the
remainder of the paper. 

A first step is to define the data sample inside which \LV-SUSY should be
searched for. To this purpose, 
a selection of triggers designed for \LV-SUSY scenarios are
proposed for ``mid-luminosity'' running of the LHC ($\Lum=3\ttn{33}\scm$) in section \ref{sec:triggers}. 

Before going further, we note that the conventional 
procedure of studying a few benchmark points in detail is insufficient in the
enlarged parameter space opened up by the $L$-violating couplings,
even more so since the presence of such couplings also remove important
cosmological contraints on the mSUGRA
parameters. In particular, the Lightest Supersymmetric
Particle (LSP) is no longer required to be neutral \cite{ellis84}, 
and essentially no bound on the mSUGRA parameters can be 
 obtained by requiring 
a relic density less than or equal to the
density of dark matter in the universe. 

In section \ref{sec:discovery}, we make 
a first attempt at coming to grips with the size of the parameter space,
proposing a method which relies 
on neural networks and the grouping of individual
scenarios into classes, whereby regions rather than individual 
points become the studied objects in parameter space. 
In this work, 
we define just 2 classes of \LV-SUSY models, each containing 15 individual
scenarios and a class consisting of 5 MSSM scenarios for reference. 
For each class, a neural network is trained with post-trigger 
events from all the scenarios in the class against background events, 
allowing the network to pick out general
qualities common to each class without over-fitting to a particular
model. Although we study only a few scenarios here, 
the usefulness of this method should become apparent when considering the
requirements posed by more comprehensive scans over parameter space.

After some remarks pertaining to the dangers of 
using neural networks, we
present results for the estimated ATLAS sensitivity, $S/\sqrt{S+B}$, for all scenarios
with an assumed integrated luminosity of 30\fb$^{-1}$. The subsequent step of
pinning down the model parameters once a signal has been observed, is clearly
more model dependent and is not covered here. An up-to-date review can be found
in \cite{allanach99}.

A brief outlook and concluding remarks are given in section \ref{sec:conc}.

\section{L-Violating SUSY \label{sec:LV}}
In this part of the paper, we do not present new results. Rather, we
summarize some theoretical considerations concerning $R$-violation which,
although well documented in the literature, may not be in the active memory
of all readers.

Despite the many attractive features of Supersymmetry, most notably that it
provides a natural solution to the hierarchy problem, it 
is well known that the most general supersymmetric Lagrangian 
(containing all terms obeying 
$SU(3)_C\times SU(2)_L\times U(1)_Y$ gauge invariance 
and Supersymmetry) is utterly incompatible with experiment, 
regardless of whether the assumption of minimal particle content is
made or not. 

The reason is that baryon and lepton number conservation is not
guaranteed by any of the symmetries just mentioned, and the accidental conservation of
these quantum numbers in the SM does not hold when extending the SM with Supersymmetry;
the full SUSY Lagrangian contains
renormalizable lepton and baryon number violating operators \cite{weinberg82} 
which are suppressed only by the SUSY
breaking mass scale squared. Unless a much more powerful suppression
mechanism is also at work, these operators 
result in a lifetime for the proton which would be measured 
in fractions of a second, to be compared with the experimental 
bound $\tau_p > 1.6\times 10^{33}\yr$ (at 90\% CL) obtained by Super-Kamiokande
in the $p\to e^+\pi^ 0$ channel \cite{shiozawa98}. 

The possibility that the couplings responsible for proton decay are 
just naturally small but not zero
is almost out of the question. Assuming $M_{SUSY}\approx
 1\TeV$, the product of the \BV\ and \LV\ couplings involved in proton decay 
is required to be less than $\tn{-25}$ \cite{dreiner98}. 
Since at least one of the couplings is then
 forced to have a value below $\tn{-12}$ without any obvious suppression
 mechanism at work, it is more natural to assume that there is some additional
 symmetry in the theory, giving zero couplings either for the $B$-violating terms,
 the $L$-violating terms, or both.

Thus, in the Minimal
Supersymmetric Standal Model (MSSM), a discrete symmetry, $R$-parity
\cite{farrar78}, is customarily imposed which ensures the conservation of
baryon as well as lepton number in the supersymmetric Lagrangian. 
Since there are strong indications 
that non-gauged symmetries are maximally violated by 
quantum gravity effects (leading to wormhole-induced proton decay
\cite{gilbert88}) whereas gauged symmetries are totally stable against such
effects \cite{krauss89}, we shall here assume 
that the proton-protecting symmetry is in fact a so-called 
\emph{discrete gauge symmetry} \cite{wegner71}.
In this case, it is
of importance to note that although $R$-parity conservation leads to a
comparatively simple phenomenology and a natural dark matter candidate (the LSP), the
problem with proton decay is not satisfactorily solved when embedding the
supersymmetric theory into more fundamental frameworks (such as GUTs) 
where baryon and lepton number violation can appear at some higher scale. 

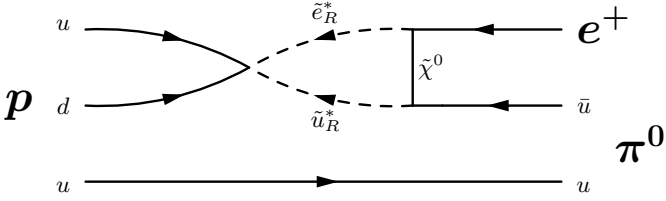
\begin{figure}
\setlength{\unitlength}{0.9pt}
\begin{center}
\begin{fmffile}{dim5}
\begin{fmfgraph*}(250,80)\fmfset{arrow_len}{3mm}
\fmfforce{0.1w,0.9h}{ll1}
\fmfforce{0.1w,0.5h}{ll2}
\fmfforce{0.1w,0.1h}{ll3}
\fmfforce{0.3w,0.9h}{l1}
\fmfforce{0.3w,0.5h}{l2}
\fmfforce{0.3w,0.1h}{l3}
\fmfforce{0.65w,0.9h}{v2u}
\fmfforce{0.65w,0.5h}{v2l}
\fmfforce{0.7w,0.9h}{r1}
\fmfforce{0.7w,0.5h}{r2}
\fmfforce{0.7w,0.1h}{r3}
\fmfforce{0.9w,0.9h}{rr1}
\fmfforce{0.9w,0.5h}{rr2}
\fmfforce{0.9w,0.1h}{rr3}
\fmf{fermion,tension=0.25,left=0.14}{ll1,v}
\fmf{fermion,tension=0.25,right=0.14}{ll2,v}
\fmf{fermion}{ll3,rr3}
\fmf{scalar,tension=0.25,right=0.15,label=\small $\ti{e}_R^*$,label.dist=5}{v2u,v}
\fmf{scalar,tension=0.25,left=0.15,label=\small $\ti{u}_R^*$,label.dist=4}{v2l,v}
\fmf{plain,tension=0.55}{r2,v2l}
\fmf{fermion}{rr2,r2}
\fmf{plain,label=\small\neut,label.dist=2,label.side=left}{v2u,v2l}
\fmf{fermion,tension=0.2}{rr1,v2u}
\fmfv{label=\LARGE\boldmath{$e^+$}}{rr1}
\fmfv{label=\small$\bar{u}$}{rr2}
\fmfv{label=\small $u$}{ll1}
\fmfv{label=\small $d$}{ll2}
\fmfv{label=\small $u$}{ll3}
\fmfv{label=\small $u$}{rr3}
\fmf{phantom,label=\LARGE\boldmath{$p$},label.dist=20}{ll1,ll3}
\fmf{phantom,label=\LARGE\boldmath{$\pi^0$},label.side=left,label.dist=20}{rr2,rr3}
\end{fmfgraph*}
\end{fmffile}
\caption{An example of proton decay proceeding 
via a 5-dimensional operator violating both $L$ and $B$ but conserving
$R$. Due to the Majorana nature of the neutralino, it is drawn without an
arrow. \label{fig:dim5}}
\end{center}
\vspace*{-\tfcapsep}\end{figure}

In an effective
theory valid around the electroweak energy scale, the interactions
mediated by super-heavy resonances associated with a higher-scale theory, 
take the form of non-renormalizable operators (i.e.\ operators of mass dimension
$d\ge5$) suppressed by $d-4$ powers of the high scale.
Such operators, violating baryon and/or lepton number 
are generally \emph{not} forbidden by $R$-parity conservation (see fig.~\ref{fig:dim5}), 
and it has been demonstrated that operators of dimension 5, 
being suppressed by only one power of the high (e.g.\ GUT) scale, will cause too 
rapid proton decay unless their couplings are suppressed by 
several orders of magnitude \cite{hinchliffe93}. This should, by itself,
provide a powerful argument for exploring alternatives to $R$-parity. 

In \cite{ibanez92}, a systematic analysis of discrete gauge symmetries was
carried out with the surprising conclusion that only $R$-parity and a $Z_3$
symmetry equivalent to $L$-conservation are anomaly-free, although some model
dependence of this analysis \cite{banks92} implies that also conservation of
baryon number could be assured by viable symmetries. 
We shall here denote the two possibilities
by $L$-parity and $B$-parity, 
though this
should \emph{not} be taken to mean that these symmetries stand for exact lepton number
conservation and exact baryon number conservation (for obvious reasons to do
with the matter-antimatter asymmetry of the universe).  
These names only reflect that in the supersymmetric part of the theory, the
corresponding quantities are conserved. They may still be violated by GUT
scale operators and will certainly be so by electroweak sphalerons. By far
the most interesting observation, however, is that both $L$-parity and
$B$-parity \emph{but not} $R$-parity can forbid the dangerous dimension 5
contributions to proton decay. 

Since a large number of current 
GUT and Planck scale theories do contain baryon and lepton number violation
exterior to the supersymmetric MSSM framework, which will result in exactly
this kind of contributions, it is important
that experiments planning to explore SUSY are well prepared for the very
distinct experimental signatures which are the hallmarks of $R$-parity
violating scenarios, most importantly the consequences of a decaying LSP,
of single sparticle production, and of the removal of the cosmological constraint
that the LSP be neutral.

In this article, we assume the framework of the MSSM with the modification 
that lepton number is violated. This leads to the MSSM interaction Lagrangian 
being enlarged by the following terms \cite{dreiner98}:
\begin{eqnarray}
\mathcal{L}_{\LV} & = & \lambda_{ijk}\left(
\bar{\nu}_{Li}^c e_{Lj}\tilde{e}_{Rk}^* +\nu_{Li}\tilde{e}_{Lj}\bar{e}_{Rk}
+ \tilde{\nu}_{Li}e_{Lj}\bar{e}_{Rk} - i\leftrightarrow j 
\right)  \nonumber \\ & & +
\lambda'_{ijk}\left( \bar{\nu}_{Li}^c d_{Lj}\tilde{d}^*_{Rk} +
\nu_{Li}\tilde{d}_{Lj}\bar{d}_{Rk} + \tilde{\nu}_{Li}d_{Lj}\bar{d}_{Rk}
\right. \nonumber\\
 & & \left.\hspace*{10mm} - \bar{e}^c_{Ri}u_{Lj}\ti{d}^*_{Rk} -
 e_{Li}\ti{u}_{Lj}\bar{d}_{Rk} - \ti{e}_{Li}u_{Lj}\bar{d}_{Rk} \right) \nonumber\\
 & & + \mbox{h.c.}
\label{eq:LVLagrangian}
\end{eqnarray}
where $i,j,k$ are generation indices (summation implied) 
and $\lambda_{ijk}$ is antisymmetric in its
first two indices. The terms in the first line of the above equation 
are customarily denoted LLE terms and the terms in second LQD, in reference
to the superfields appearing in the superpotential from which
eq.~(\ref{eq:LVLagrangian}) is derived. 
All of these terms contain two (SM) fermions and one (SUSY) scalar. 
The oddness under $R$ is thus directly visible here, since the SM fields have $R=+1$
and the SUSY fields $R=-1$. Also,
it is clear that each sfermion can decay in a number of ways to two SM fermions
via these couplings. $L$-violating neutralino, chargino, and gluino decays
are forced to proceed via one or more intermediate scalar resonances. 

A complete list of decay modes and full matrix elements for all decays of
MSSM sparticles to SM particles can be found in 
\cite{dreiner00}. Of these, we have
implemented all but the gluino decays 
 into the \pythia\ event generator, publically available from version
6.2, with documentation included in the recently revised \pythia\
manual \cite{pythia6.2}. The conclusions of the present paper are not
expected to be changed significantly 
by the inclusion of \LV\ gluino decays since the gluino is
typically heavy and thus has a number of other, unsuppressed decay channels
available. Details about the \pythia\ implementation can be found 
in \cite{skands_thesis}. 

Note that the sfermion decay matrix elements in \cite{dreiner00} have been
  checked analytically by the author, and that the matrix elements listed
  there are not directly applicable in \pythia. Both \pythia\ and
  \herwig\ (\isawig) follow \cite{haber85,gunion86} for the chargino and neutralino
  mixing conventions, 
  but \herwig\ uses the opposite convention \cite{baer94} for the
  sfermion mixing angles, yielding a relative transposition between the two
  programs. In addition, extensive counter-checks 
  were made between the two programs to make sure they agree\footnote{This
  resulted in a few bug-fixes also in the \herwig\ code, so \isawig\ has
  had a few bugs in the ME calculation up to and including v.1104.}. Some
  non-negligible differences exist:
\begin{enumerate}
\item \isawig\ does not use running masses in the evaluation of higgsino-type
  couplings whereas \pythia\ does. This has been observed to lead to
  substantial differences between the widths calculated by the two
  programs (in rare cases factors of 2 or more).
\item $\alpha_s$ at $M_Z$ is used by \isawig\ in the calculation of the gluino decays. In the
  forthcoming update of \pythia\ where gluino decays are included, $\alpha_s$
  will be evaluated at the mass of the gluino.
\item \pythia\ and \herwig\ (\isasusy) do not use the same RGE's, so many
  parameters can differ quite substantially at the EW scale from the same GUT
  input, creating an ``artificial'' dissimilarity between the two
  programs. Ideally, one should compare models with identical EW scale 
  mass spectra.
\end{enumerate}
\section{mSUGRA Models \label{sec:mSUGRA}}
The models used in this study have not been chosen among the ones 
initially suggested in the ATLAS Physics TDR \cite{atlastdr}. 
This is partly due to the
exclusion of most of these points by LEP (essentially from bounds on the Higgs
mass), and partly since it is interesting to enable a direct comparison
between the capabilities of the LHC and other, future experiments. 
The 5 mSUGRA points shown in table \ref{tab:sugrapoints}
have therefore been selected among 14 points which were defined by the CLIC
physics study group \cite{deroeck01}. More recently - too late to be included
in the studies reported here - a new set of 
standard benchmarks points were proposed \cite{ellisbm}
which have now been adopted by both the CLIC and other linear collider
communities. 
See \cite{ellisbm} for mass spectra analogous to table
\ref{tab:sugrapoints}. To aid comparison, we briefly list the
main differences/similiarities between our points and the new benchmark
points. 
\begin{itemize}
\item Point 'A' is a (150\GeV) 
lighter version of $P_2$. The total SUSY (pair production) cross section is
approximately 4 times larger than for $P_2$: $\sigma_A \approx 4\sigma_{P_2}$.
\item Points 'B', 'C', 'I', and particularly 'G' are similar to
  $P_9$. $\sigma_{P_9} \approx 0.5\sigma_B \approx 5\sigma_C \approx
  2\sigma_I \approx 3.5\sigma_G$. 
\item Point 'D' is dissimilar to all our points. It is closest to $P_2$, but
  has a 500\GeV\ lighter gluino, and 400\GeV\ lighter squarks which increases
  the LHC cross section by an order of magnitude.
\item Point 'E' is a lower-mass version of $F_2$, point 'F' a higher-mass
  version. $\sigma_{F_2}\approx13\sigma_F \approx 0.05\sigma_E$.
\item Point 'H' is similar to $P_7$ but slightly heavier, giving a total LHC
   cross section\footnote{Point 'H' gave a $\ti{\tau}^+_1$ LSP when using the
     GUT input parameters in \cite{ellisbm} with ISASUGRA. 
     We therefore chose a 30\GeV\ larger value for $m_0$, retrieving $\neut_1$
     as the LSP.} $\sigma_{H} \approx \textstyle\frac14\sigma_{P_7}$.
\item Point 'J' is a high-$\tan\beta$ version of $P_2$ with very close to 
  identical total cross section. The mass hierarchies are also very similar.
\item Point 'K' is a large-$m_{1/2}$ sister of the only large-$\tan\beta$
  model included in this work, $P_{12}$. The correspondingly larger masses
  yield a cross section $\sigma_{K} \approx 0.04\sigma_{P_{12}}$. The mass
  spectrum is similar to that of $P_7$.
\item We have not included points at $\tan\beta=45$, but point 'M' can be
  seen as an even heavier version of point 'K'$\approx P_{12}$ with a total 
  cross section of only 0.1 femtobarns, and point 'L'
  has a mass spectrum not greatly different from $P_9$ but with heavier
  squarks and gluinos giving a cross section an order of magnitude lower than
  for $P_9$.
\end{itemize}
The total cross sections compared here 
include all MSSM pair production. Single
sparticle production will give additional contributions depending on the
strengths of the $R$-violating couplings.

One should keep in mind that all these points are
defined for the MSSM, and as such have neutralino LSP's, a property 
which it has already been mentioned is not necessary in \RV-SUSY
scenarios. 
\begin{table}[t]
\begin{center}
{\small
\begin{tabular}{crrrrr}\toprule
 & \hspace*{0.4cm}$\mathbf{P_2}$ & \hspace*{0.4cm}$\mathbf{P_7}$ & \hspace*{0.4cm}$\mathbf{P_9}$ & \hspace*{0.4cm}$\mathbf{P_{12}}$ & \hspace*{0.4cm}$\mathbf{F_2}$ \\ \cmidrule{1-6}
\multicolumn{6}{c}{\textbf{GUT Parameters}}\\ 
$\tan\beta$     &   5 &  10 &  20 &  35 &   10 \\
$m_0$           & 170 & 335 & 100 &1000 & 2100 \\
$m_\frac12$     & 780 &1300 & 300 & 700 &  600 \\
sign$(\mu)$     &   + &   + &   + & $-$ &    + \\
$A_0$           &   0 &   0 &   0 &   0 &    0 \\
\cmidrule{1-6}\multicolumn{6}{c}{\textbf{Mass Spectrum}}\\
$h^0$           & 118 & 123 & 115 & 120 & 119 \\
$A^0, H^\pm, H^0$&1100 &1663 & 416 & 944 &2125 \\\cmidrule{1-6}
$\neut_1$       & 325 & 554 & 118 & 293 & 239 \\
$\neut_2,\charg_1$&604&1025 & 217 & 543 & 331 \\
$\neut_3$       & 947 &1416 & 399 & 754 & 348 \\
$\neut_4,\charg_2$&960&1425 & 416 & 767 & 502 \\\cmidrule{1-6}
$\ti{g}$        &1706 &2752 & 707 &1592 &1442 \\\cmidrule{1-6}
$\ti{e}_R,\ti{\mu}_R$&336&584&156 &1031 &2108 \\
$\ti{\tau}_1$   & 334 & 574 & 126 & 916 &2090 \\
$\ti{e}_L,\ti{\mu}_L$&546&917&231 &1098 &2126 \\
$\ti{\tau}_2$   & 546 & 915 & 240 &1051 &2118 \\
$\ti{\nu}$      & 541 & 913 & 217 &1095 &2125 \\\cmidrule{1-6}
$\ti{q}_R$      &1453 &2333 & 612 &1612 &2328 \\
$\ti{b}_1$      &1403 &2262 & 566 &1412 &2010 \\
$\ti{t}_1$      &1189 &1948 & 471 &1241 &1592 \\
$\ti{q}_L$      &1514 &2425 & 633 &1663 &2343 \\
$\ti{b}_2$      &1445 &2312 & 615 &1482 &2310 \\
$\ti{t}_2$      &1443 &2286 & 648 &1451 &2018 \\ 
\cmidrule{1-6}\multicolumn{6}{c}{\textbf{LHC Parameters}}\\
 & \hspace*{0.4cm}$\mathbf{P_2}$ & \hspace*{0.4cm}$\mathbf{P_7}$ & \hspace*{0.4cm}$\mathbf{P_9}$ & \hspace*{0.4cm}$\mathbf{P_{12}}$ & \hspace*{0.4cm}$\mathbf{F_2}$ \\
$\sigma_{\mathrm{SUSY}}$ [fb]&130&3.9&24000&110&110\\
$\times 30\mathrm{fb}^{-1}$ &3900&114&720000&3300&3300\\
\bottomrule
\end{tabular}}
\caption[\small Selected points in the mSUGRA space]{Selected points of
analysis in the mSUGRA parameter space, mass spectra as obtained with
\isasusy, and total SUSY pair production cross section at the
LHC.\label{tab:sugrapoints}} 
\end{center}                 
\vspace*{-\tfcapsep}\end{table}
Though we shall not do so here, it is certainly advisable to explore the
experimental consequences of non-neutralino LSP's in more detail.

Both our points and the points in \cite{ellisbm} 
assume a vanishing trilinear coupling at the GUT scale, i.e.\ 
$A_0=0$. In connection with this work, a small study of the direct consequences of
that assumption upon the results presented here was performed. Varying $A_0$
between 0 and 500\GeV\ for $P_2$, $P_9$, and $P_{12}$ gave only a weak
variation ($\mathcal{O}(5\%)$) of the 
semi-inclusive $L$-violating 
branching ratios (e.g.\ $BR(\neut_1\to qq\nu)$), and so apart from the
consequences of the 
distortion of the mass spectrum caused by $A_0\neq 0$, the main
signatures (number of leptons, number of jets, etc.) should be only mildly
affected by changes to this parameter. 

In addition to the mSUGRA parameters come the 9 $\lambda$ (LLE) and the 27 $\lambda'$
(LQD) couplings for which the scenarios listed in table \ref{tab:lambdapoints} have
been studied for each mSUGRA point. The reason we do not consider
couplings
larger than $10^{-2}$ is partly due to the present experimental 
bounds \cite{rvbounds} and partly due to the fact that some implicit
approximations in this analysis would break down for larger
couplings: 1) the \RV\ couplings are not included in the RGE
evolution of the SUSY parameters and 2) single sparticle production is not
simulated. Note that the large-coupling ($a$)
scenarios only observe the limits on individual couplings given 
in \cite{rvbounds}, 
not the limits on products of the couplings. Thus, these scenarios are useful
for studying the consequences of having many large couplings, but one should
keep in mind that they are unrealistic in that not \emph{all} of the
couplings can be simultaneously large. 

On the other hand, if the \RV\  
couplings get significantly smaller than $10^{-4}$, the LSP lifetime can
become so large that it decays outside the detector, 
mimicking the $R$-conserving
scenarios which have already been extensively studied. 
For example, for the mSUGRA point $F_2$, 
setting all $\lambda$ couplings to
$10^{-6}$ and all $\lambda'$ couplings to zero 
results in a decay length for the LSP of $\tau c = 40\ \!\mbox{m}$. In the
intermediate range, one may see the LSP decay directly
inside the fiducial volume of the detector (see e.g.\ \cite{dreiner91}), yet 
we abstain from relying on such a spectacular signature here so as not to be
overly optimistic in our results. 
\begin{table}[tb]
\center\small \begin{tabular}{cccc}\toprule
& \multicolumn{3}{c}{\boldmath\bf\LV\ \ \ C\ O\ U\ P\ L\ I\ N\ G\ \ \ M\ O\ D\ E\ L\ S}\\
& $a$ & $b$ & $n$
\\ \cmidrule{1-4} \multicolumn{4}{c}{L\ L\ E} 
\\
        $\begin{array}{c}
        \lambda_{ijk} \\ 
        \lambda'_{ijk}
        \end{array}$ &
        $\begin{array}{c}
        \hspace*{2mm}10^{-2}\hspace*{2mm} \\ 0 
        \end{array}$ &
        $\begin{array}{c}
        \hspace*{2mm}10^{-4}\hspace*{2mm} \\  0 
        \end{array}$ &
        $\begin{array}{c}
        \sqrt{\hat{m}_{e_i}\hat{m}_{e_j}\hat{m}_{e_k}} \\ 
        0
        \end{array}$
\\ \cmidrule{1-4} \multicolumn{4}{c}{L\ Q\ D} 
\\
        $\begin{array}{c}
        \lambda_{ijk} \\ 
        \lambda'_{ijk}
        \end{array}$ &
        $\begin{array}{c}
         0 \\ 10^{-2} 
        \end{array}$ &
        $\begin{array}{c}
        0 \\ 10^{-4} 
        \end{array}$ &
        $\begin{array}{c}
        0 \\ 
        \sqrt{\hat{m}_{e_i}\hat{m}_{q_j}\hat{m}_{d_k}}
        \end{array}$
\\ \cmidrule{1-4} \multicolumn{4}{c}{L\ L\ E\ \ \ +\ \ \ L\ Q\ D}
\\
        $\begin{array}{c}
        \lambda_{ijk} \\ 
        \lambda'_{ijk}
        \end{array}$ &
        $\begin{array}{c}
        10^{-2} \\ 10^{-2} 
        \end{array}$ &
        $\begin{array}{c}
         10^{-4} \\  10^{-4} 
        \end{array}$ &
        $\begin{array}{c}
        \sqrt{\hat{m}_{e_i}\hat{m}_{e_j}\hat{m}_{e_k}} \\ 
        \sqrt{\hat{m}_{e_i}\hat{m}_{q_j}\hat{m}_{d_k}}
        \end{array}$
\\ \bottomrule
\end{tabular}
\caption[\small Selected points in the $\lambda-\lambda'$ space]{Selected
        points of analysis in the $\lambda-\lambda'$ parameter
        space. The last column corresponds to the ``natural coupling''
        scenario proposed in \cite{hinchliffe93}. $\hat{m}\equiv \frac{m}{v}
        = \frac{m}{126\GeV}$ and $m_{q_j} \equiv
        {\textstyle\frac12}(m_{u_j}+m_{d_j})$. These models 
 (coloumn $n$)
will be referred to as ``nLLE'', ``nLQD'', and ``nLLE + nLQD'' 
in the text below. 
\label{tab:lambdapoints}}
\vspace*{-\tfcapsep}\end{table}
\section{Triggers for \LV-SUSY \label{sec:triggers}} 
A reasonable aim for the total \LV-SUSY dedicated trigger rate is about 1Hz. 
We here focus on rates \emph{after} the events have been filtered through the
trigger system, i.e.\ we make no distinction between trigger levels. 
This is a technical issue which requires more detailed knowledge of
the detector performance at mid-luminosity than is currently parametrized in
\atlfast. Specifically, no parametrization of the effects of pile-up at
mid-luminosity is included,
and so we here adopt a best-guess approach, 
performing the simulation without pile-up, and then
multiplying the resulting trigger rates by a factor of 5/3 to estimate the true rates 
at $\Lum=3\ttn{33}\scm$. This factor is based on 
the scaling exhibited by the inclusive electron, electron/photon, and
\ET+2jets trigger rates presented in \cite[chp.11]{atlastdr} from
low to high luminosity. 

To retain as much generality as possible in the trigger definitions, it is
sensible to use the information contained in the \LV\ superpotential terms
rather than a selection of decay modes to define the trigger menus. 
The $\lambda$ (LLE) couplings are purely leptonic, and thus single sparticle
production via these terms is not relevant for the LHC. Thus
we are searching for at least two hard leptons (which, however, may be
tauons), in most cases accompanied by \ET\ from escaping neutrinos. 
For the $\lambda'$ (LQD)
terms, we expect at least two leptonic objects accompanied by hard
jets when the dominant production mechanism is MSSM pair production. 
Note that pure jet signatures would also be possible 
through single slepton production, $q_1\bar{q}_2 \to \ti{\ell} \to
q_3\bar{q}_4$. Single squark production via LQD requries a lepton in the
initial state and is thus suppressed at the LHC.

The triggers so far investigated are listed in table
\ref{tab:triggerrates}, where background rates and efficiency ranges for all
models investigated are shown. For comparison, results for MSSM scenarios are
also given. 
\begin{table*}[t]
\begin{center}
\begin{tabular}{lcrrr}
\toprule
Trigger & \begin{tabular}{c}Background\\Rate\end{tabular} & 
\begin{tabular}{c}MSSM\\Efficiency\end{tabular} &
\begin{tabular}{c}LLE\\Efficiency\end{tabular}&\begin{tabular}{c}LQD\\Efficiency\end{tabular}\\
\cmidrule{1-5}
mu45I + mu45I& 0.2 Hz & 1 --- 5 \% & 10 --- 40 \% & 1 --- 10 \% \\ 
e45I + e45I& 0.1 Hz & 1 --- 5 \% & 1 --- 35 \% & 1 --- 10 \% \\ 
mu15I + e15I    & 0.1 Hz & 2 --- 5 \% & 20 --- 60 \% & 2 --- 15 \% \\ 
mu40I + me75    & 0.3 Hz & 10 --- 25 \% & 40 --- 75 \% & 10 --- 35 \% \\ 
e40I + me75    & 0.2 Hz & 10 --- 20 \% & 15 --- 70 \% & 10 --- 35 \% \\ 
j100 + mu40I   & 0.5 Hz & 10 --- 20 \% & 45 --- 70 \% & 10 --- 40 \% \\ 
j100 + e40I    & 0.5 Hz & 5 --- 15 \%  & 15 --- 65 \% & 10 --- 35 \% \\ 
j100 + me175   & 0.3 Hz & 50 --- 80 \% & 35 --- 80 \% & 25 --- 80 \% \\ 
3j50 + mu20I   & 0.1 Hz & 5 --- 15 \% & 45 --- 60 \% & 12 --- 40 \% \\ 
3j50 + e30I    & 0.1 Hz & 5 --- 10 \% & 15 --- 55 \% & 10 --- 35 \% \\ 
3j75 + me125   & 0.1 Hz & 30 --- 65 \% & 30 --- 70 \% & 40 --- 90 \% \\ 
\cmidrule{1-5}
Total Rate&   2.1 Hz &
60 --- 90 \% & 90 --- 99.9 \% & 60 --- 96 \%  
\\\multicolumn{4}{l}{/Combined Efficiency}
\\
\bottomrule
\end{tabular}
\caption[\small Trigger rates and efficiencies for $\Lum=3\times 10^{33}\mathrm{cm}^{-2}\mathrm{s}^{-1}$.]{Estimated trigger rates for background
processes and trigger efficiency ranges for the various MSSM points and \LV-SUSY
  scenarios studied for $\Lum=3\times
  10^{33}\mathrm{cm}^{-2}\mathrm{s}^{-1}$. The nomenclature follows the ATLAS
  standard, where e.g.\ 
  ``mu45I'' means an isolated muon with $p_T > 45\GeV$, ``me'' stands for
  missing (transverse) energy, and ``3j50'' means 3 jets, each with
  $p_T>50\GeV$. The total rate is smaller than the sum of the individual
  rates since there is a certain overlap, and the combined efficiencies 
  can be larger than the individual
  efficiencies, since there is not a \emph{total} overlap between the triggers. \label{tab:triggerrates}}
\end{center}
\vspace*{-\tfcapsep}\end{table*}
\begin{table}
\begin{tabular}{lrrr}\toprule
\multicolumn{4}{c}{EVENT RATES AND SAMPLE SIZES}
\\\cmidrule{1-4}
\textbf{Process} & \boldmath\bf$\sigma$ [mb] &\bf\boldmath Rate
[Hz]&\boldmath\bf$N_{gen}$ \\
QCD $2\to 2$\\
\hspace*{2mm}{\scriptsize$p_T=1-10\GeV$}
& 55& 1.6\ttn{8}& 2.5\ttn{8} \\ 
\hspace*{2mm}{\scriptsize$p_T=10-75\GeV$}
& 12 & 3.7\ttn{7}& 2.2\ttn{8}  \\
\hspace*{2mm}{\scriptsize$p_T=75-150\GeV$}
& 5.5 \ttn{-3} & 1.7\ttn{4}& 1.4\ttn{7} \\
\hspace*{2mm}{\scriptsize$p_T>150\GeV$}
& 2.9\ttn{-4}& 8.7\ttn{2}& 1.1\ttn{7}\\\cmidrule{1-4} 
$t\bar{t}$ & 6.2\ttn{-7}& 1.9 &5.9\ttn{6}
\\
$Z/W$ & 1.2\ttn{-3}& 3.6\ttn{3}&1.8\ttn{8}
\\
$ZZ/ZW/WW$ & 1.2\ttn{-7}& 0.36 &5.9\ttn{6}\\
\bottomrule
\end{tabular}
\caption[\small Numbers of generated events for the trigger study]{Numbers of
generated events for the trigger study. The rates listed are total rates
before trigger for $\Lum=3\ttn{33}\scm$. QCD events with the $p_T$ of the hard
interaction below 1\GeV\ were not simulated.
\label{tab:ngentrig}} 
\end{table}
The event generation 
was performed with an augmented version of \pythia\ 6.155 \cite{sjostrand94} and
\atlfast\ v.2.53 \cite{atlfast2.0}. With respect to \atlfast, an attempt was
made at obtaining more believable muon and electron reconstruction
efficiencies by including by hand a muon efficiency of 95\% independent of
muon momentum and an electron reconstruction efficiency of 80\% for electrons
with $p_T>50\GeV$ and 70\% for electrons with $p_T<50\GeV$. These estimates
are based on the ATLAS Physics TDR \cite{atlastdr}.
 
Background cross sections, pre-trigger event rates, and
sample sizes are listed in table \ref{tab:ngentrig}. Though low-$p_T$ QCD
events have essentially no chance to pass triggers and much less the
subsequent analysis cuts, the very high
purities required for SUSY signal extraction do not immediately admit these
events to be discounted entirely. Rather, a substantial sample of
such events was generated with the object of placing an upper bound on the
number of low-$p_T$ QCD events remaining after cuts using Poisson
statistics. This will be discussed in section \ref{sec:discovery}, however some
cautiousness should be employed in interpreting the bounds obtained, since
the Monte Carlo is here being stretched far into the tails of its
$p_T $ distributions. 

We do not consider triple gauge boson production. 
The cross section (excluding Higgs-induced
production) is around 100\fb\ \cite{altarelli00}, i.e.\ 
at the same level as interesting SUSY cross sections, yet 
invariant mass cuts on pairs of jets would presumably be able to reduce this
background considerably, and so we do not believe
that these processes are dangerous as background sources.

More detailed remarks and a large selection of plots of trigger rates and
efficiencies versus thresholds can be found in \cite{skands_thesis}. 

From the efficiencies in table \ref{tab:triggerrates}, one easily sees how much
cleaner the signatures of the purely leptonic (LLE) coupling are 
compared to the signatures involving quarks (LQD) where higher thresholds,
due to the hadronic environment, mean smaller efficiencies. From this we
conclude that it would
be of interest to extend the LQD study, examining whether 2-jet triggers
and/or 3-object triggers could enhance the efficiency.

Lastly, though the trigger proposals given here are designed explicitly with
\LV-SUSY in mind, they show a certain overlap with triggers proposed for more
conventional physics. The di-muon and 3 jets + lepton triggers, 
for example, have also been
proposed for various Higgs searches. The di-electron
trigger as well as 3 jets + electron are proposed to catch $t\bar{t}$
decays. Finally, the conventional SUSY searches also make use of
multi-lepton, jets + \ET, and multi-jet signatures \cite{bystricky96}. It is
therefore not unlikely that the triggers proposed here can be incorporated
to some extent into the conventional trigger programme. 

\section{Discovery of \LV-SUSY at the LHC \label{sec:discovery}}
The main purpose here is to deliver an impression 
of what kind of signal strengths can be achieved at the LHC with 30fb$^{-1}$ integrated
luminosity. To maintain generality in view of the more than thousand
$L$-violating channels possible, we do not 
discuss invariant mass reconstruction or measurements of the SUSY parameters
in general. We focus entirely on the isolation of candidate events through
inclusive and kinematic event-shape variables in the attempt to obtain a
statistically significant signal as compared with the background expectation
after cuts. We do this for several models simultaneously (using neural
networks in the final step), an approach which is complementary to the
conventional one, where specific scenarios are studied one by one. 

Note that no attempt
is made, beyond a crude worst-case estimate, 
to include the effects of pile-up in this analysis.
\subsection{Missing Transverse Energy}
\begin{figure*}
\begin{center}
\begin{tabular}{cc}
\includegraphics*[scale=0.35]{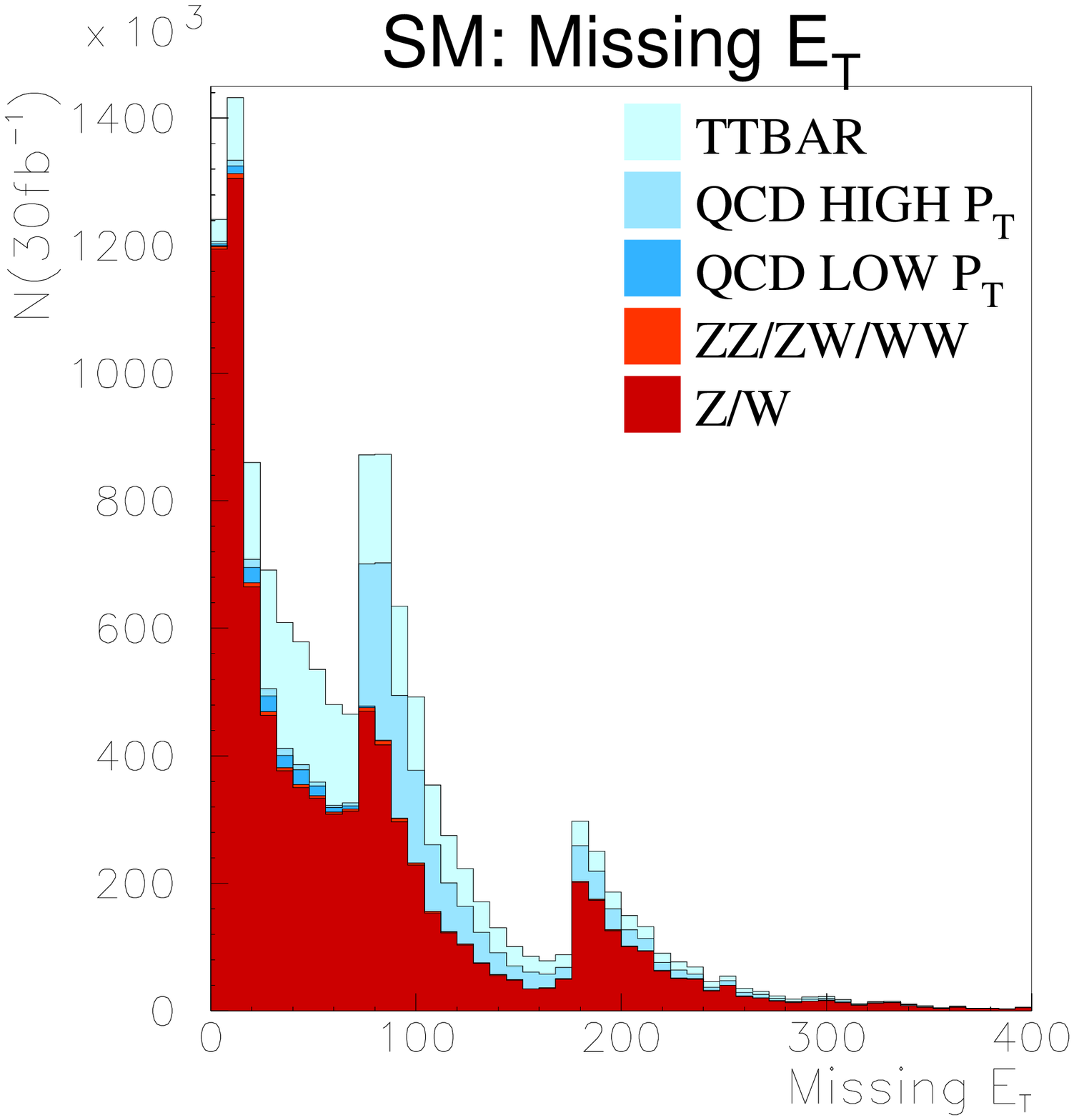}&\hspace*{-.7cm}
\includegraphics*[scale=0.35]{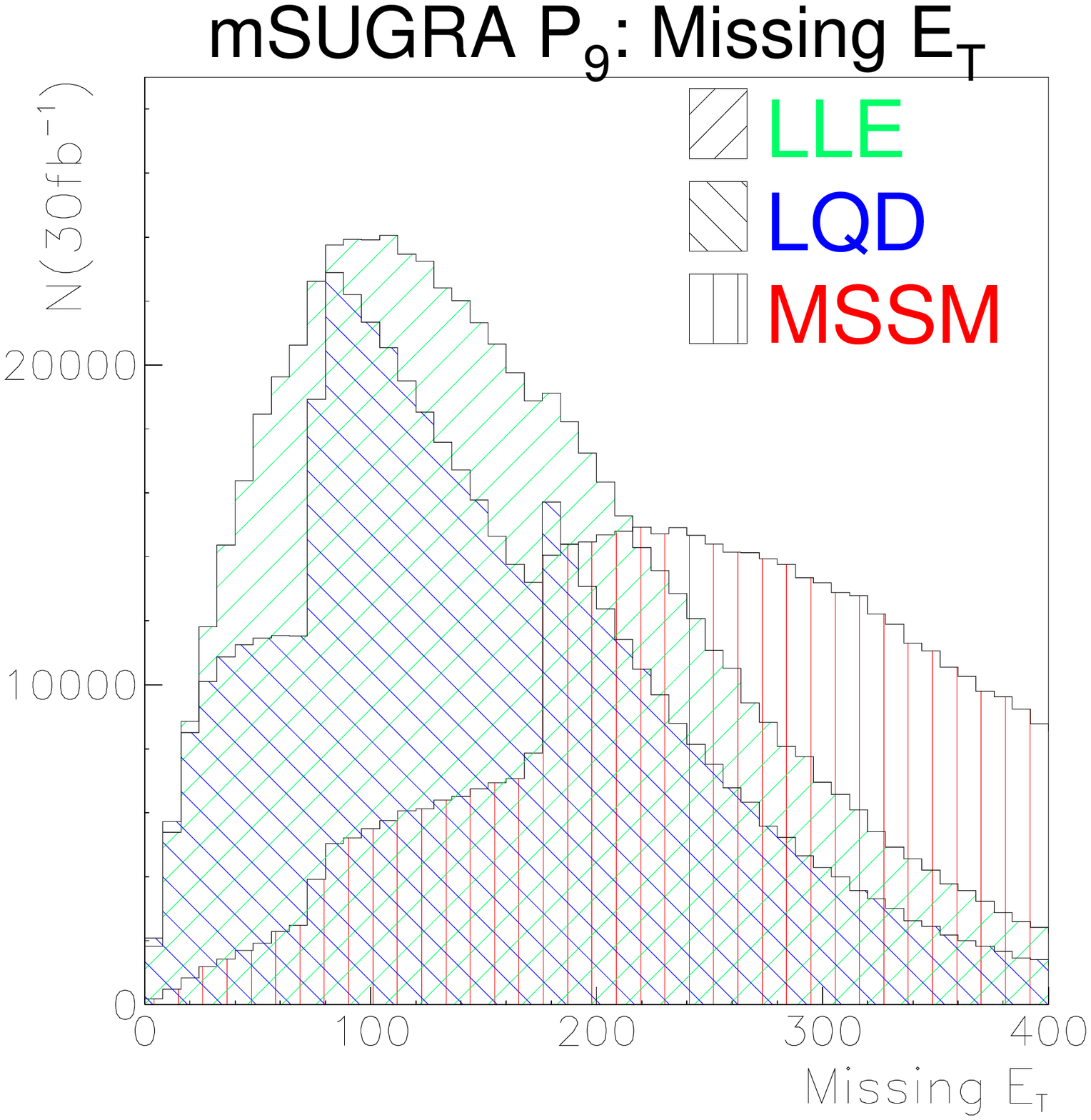}\vspace*{-0.5cm}\\
a) & b) \end{tabular}\vspace*{-8mm}\\
\includegraphics*[scale=0.7]{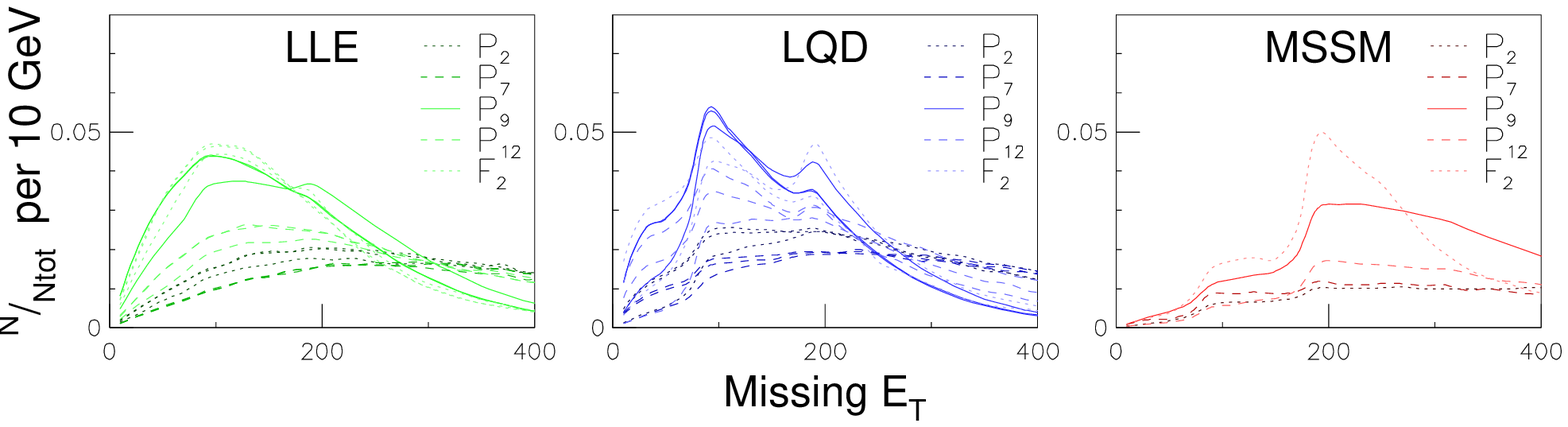}\vspace*{-.7cm}\\
c)\vspace*{-4mm}
\end{center}
\caption{a) and b): \ET\ signatures for the
SM and mSUGRA point 9$a$ models (i.e.\ all relevant \RV\ couplings set to
\tn{-2}) normalized to
30fb$^{-1}$ of data taking. ``QCD LOW $p_T$'' means events from the
$100\GeV<p_T<150\GeV$ sample and ``QCD HIGH $p_T$'' events from the
$p_T>150\GeV$ sample. 
c): Event
distributions normalized to unit area for LLE, LQD, and the MSSM for all
mSUGRA and \LV\ coupling points studied. The last row of plots is not
intended to give detailed information, merely to illustrate the spread
between the models.
\label{fig:ET}}
\end{figure*}

The post-trigger \ET\ distribution for the SM and its composition is shown in figure
\ref{fig:ET}a. Note that there are so few double gauge boson events that they are
hardly visible on the plot. The peaks at $\ET=75\GeV$ and at $\ET=175$
are due to the me75 and me175 triggers becoming active. 

In figure \ref{fig:ET}b the distributions for the most ``low-mass'' mSUGRA
 point, $P_9$, are shown for $P_{9}$(MSSM), $P_{9a}$(LLE), and
 $P_{9a}$(LQD). The degradation of the \ET\ signature in the \LV\ models is
 evident, though one observes by comparing with fig.~\ref{fig:ET}a that a
 cut at low \ET\ values is still possible. Note that the \ET\ trigger peaks
 mentioned above are absent for the LLE scenario, this simply because the LLE
 scenarios do not rely to so great an extent on the \ET\ triggers, c.f.\
 table \ref{tab:triggerrates}.
 
The full range of mSUGRA models
 are plotted in figure \ref{fig:ET}c. To enable the models to be
 shown on the same scale and to be distinguished from each other, 
 the histograms have been normalized to unit area and smoothed. These
 plots are not intended to give detailed information, but rather to
 illustrate the spread between the models. Models with heavy squarks and
 gluinos ($P_7$, $F_2$, and to some extent $P_{12}$) have rather flat, 
MSSM-like signatures whereas the models with 
lighter sparticles are peaked at low \ET. This is an important point, since
 the light-sparticle scenarios have high production cross sections and the
 heavy-sparticle ones low cross sections. 
A cut on $\ET>100\GeV$ seems a reasonable compromise between losing events
 in the peaked distributions (where we have many events anyway) and efficient
 background rejection required for the heavier scenarios (where we lose
 little by the cut but have fewer events).
\begin{figure}
\begin{center}
\includegraphics*[scale=0.2]{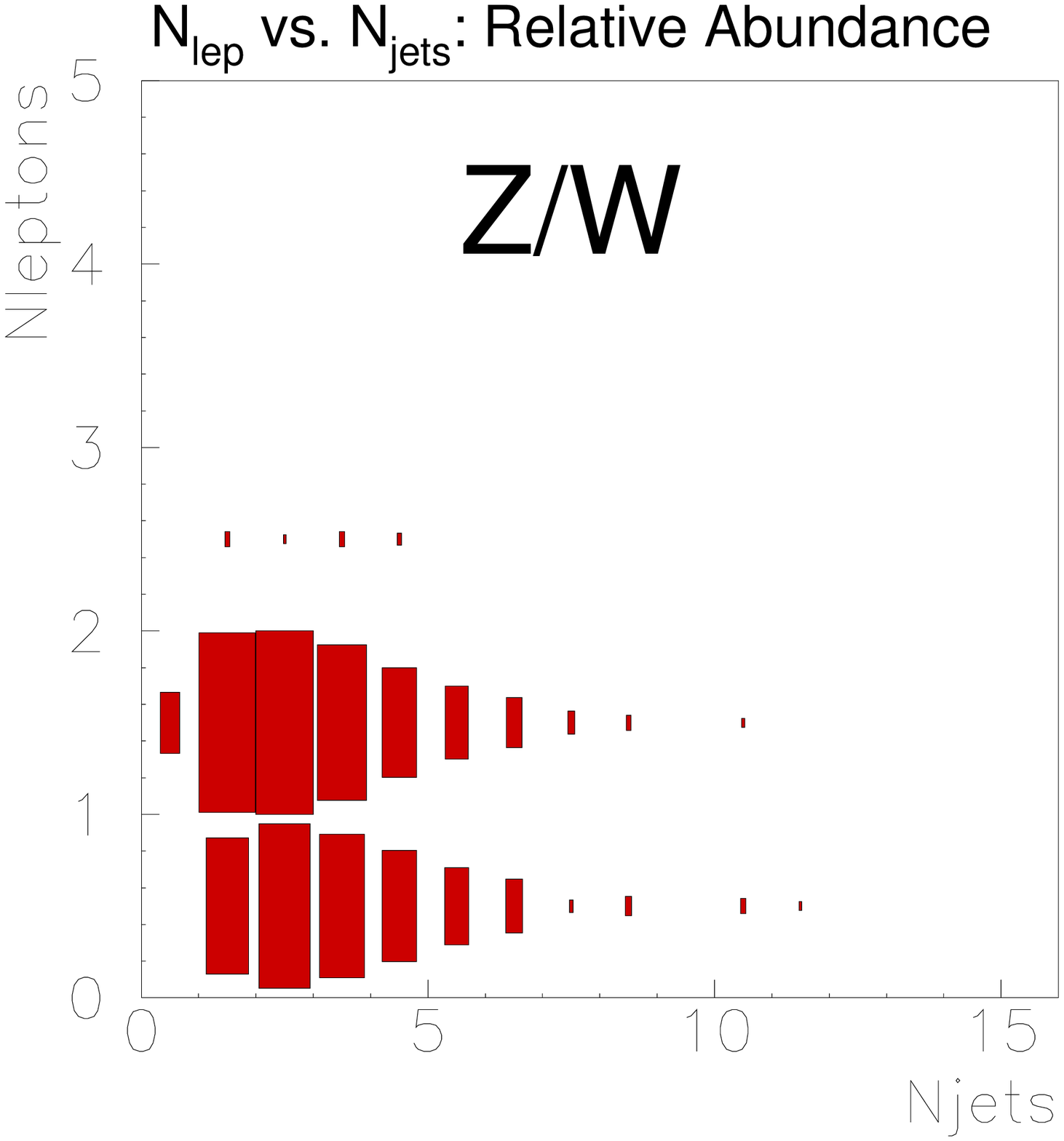}
\includegraphics*[scale=0.2]{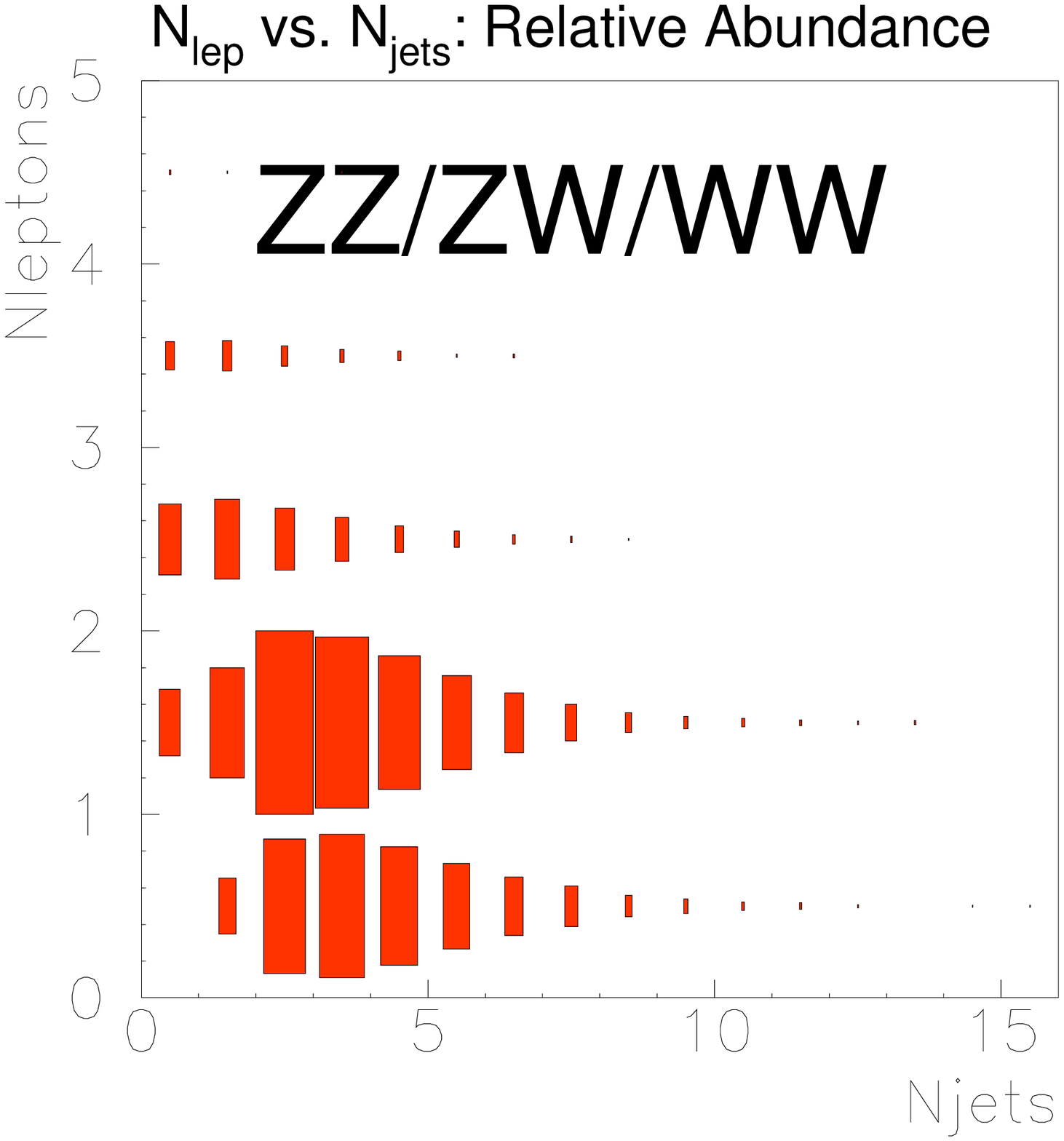}\vspace*{0.5mm}\\
\includegraphics*[scale=0.2]{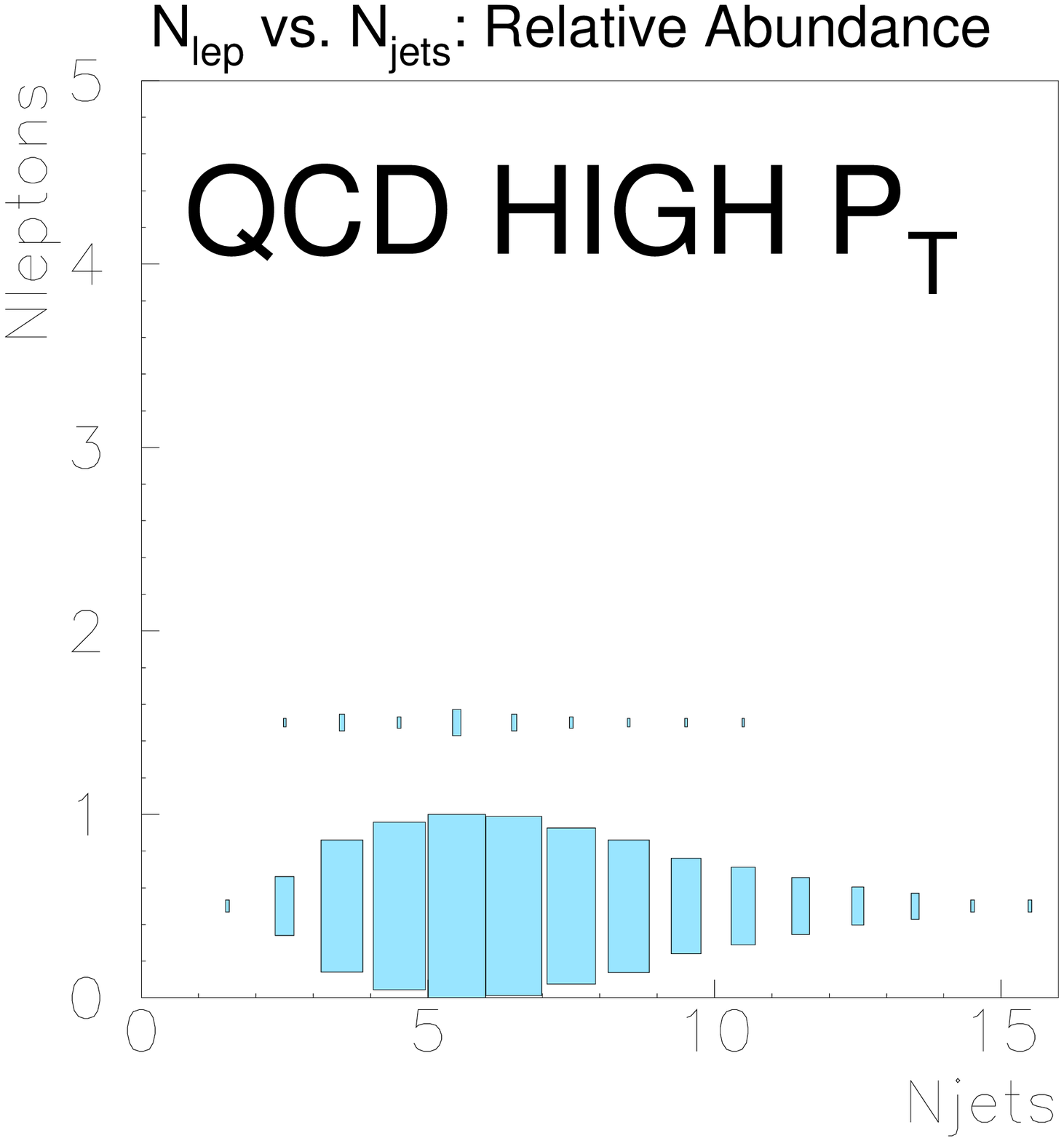}
\includegraphics*[scale=0.2]{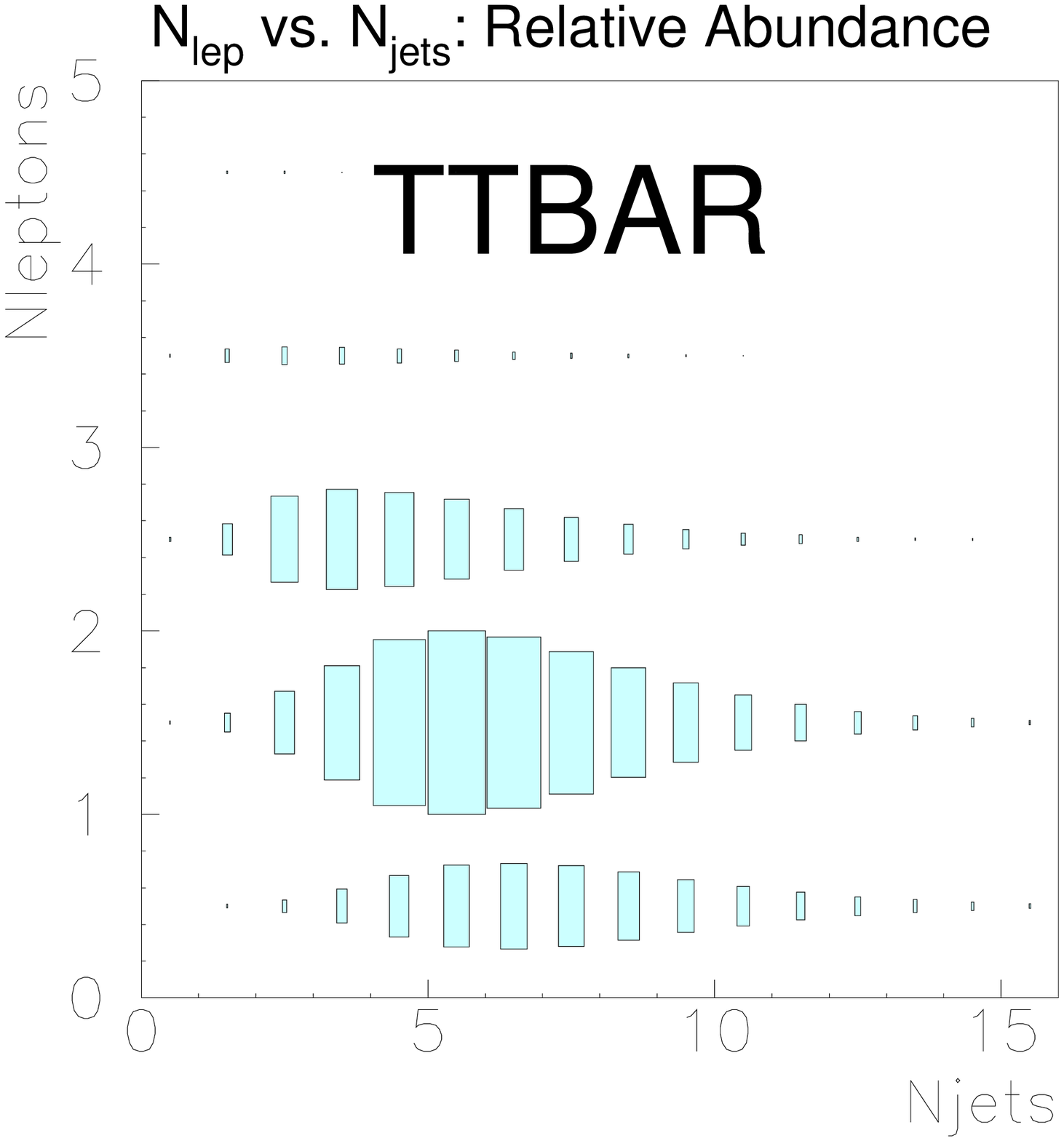}
\vspace*{-3mm}\end{center}
\caption{
Lepton versus jet multiplicity (see text) for background events 
surviving the cut on \ET. The numbering of the bins is such that
the events with 0 jets are in the
bin to the right of the number 0 and events with 0 leptons are in the bin
above the number 0. 
\label{fig:jl_sm}}
\end{figure}
\begin{figure}
\begin{center}
\includegraphics*[scale=0.2]{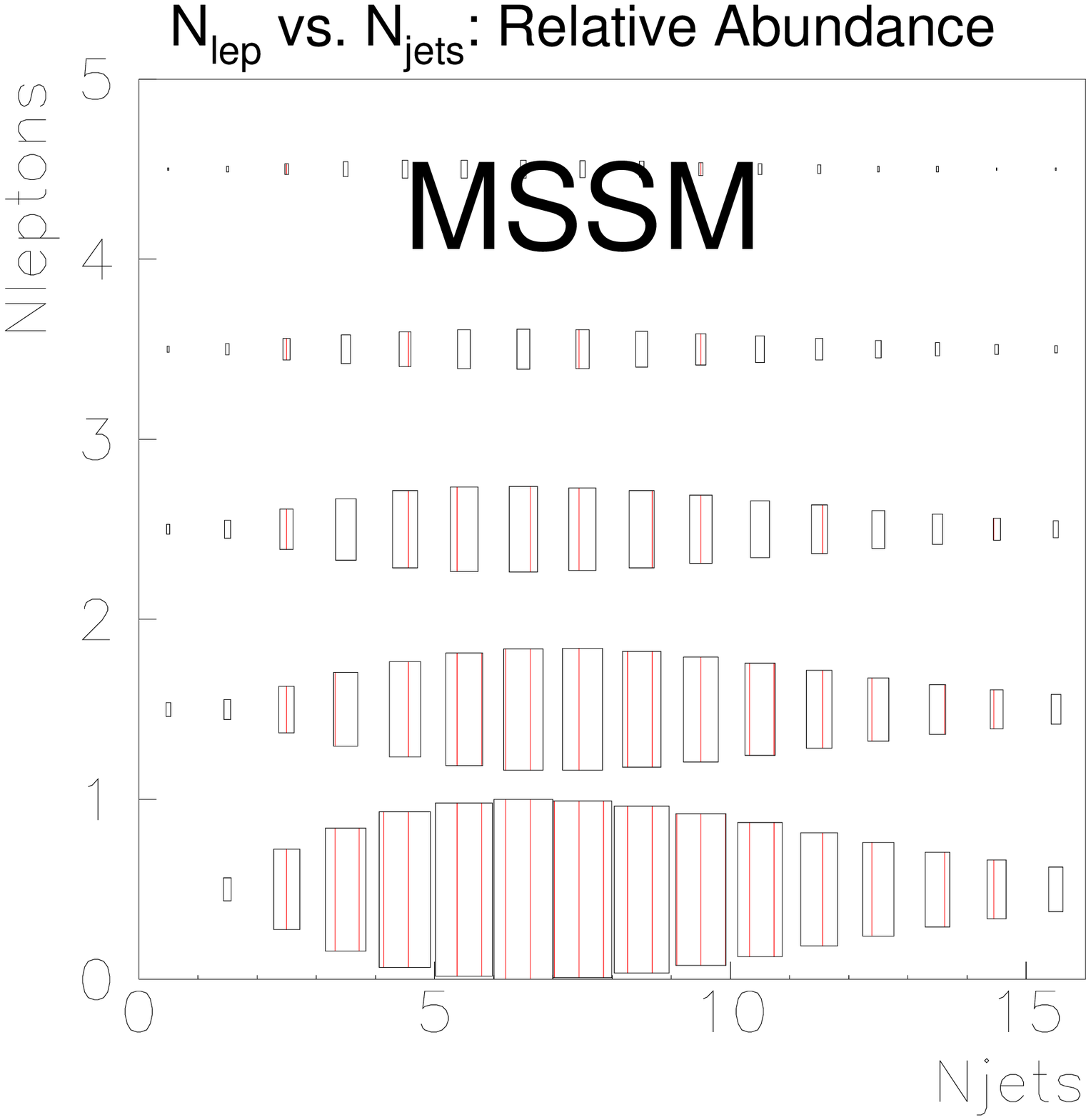}\vspace*{0.5mm}\\
\includegraphics*[scale=0.2]{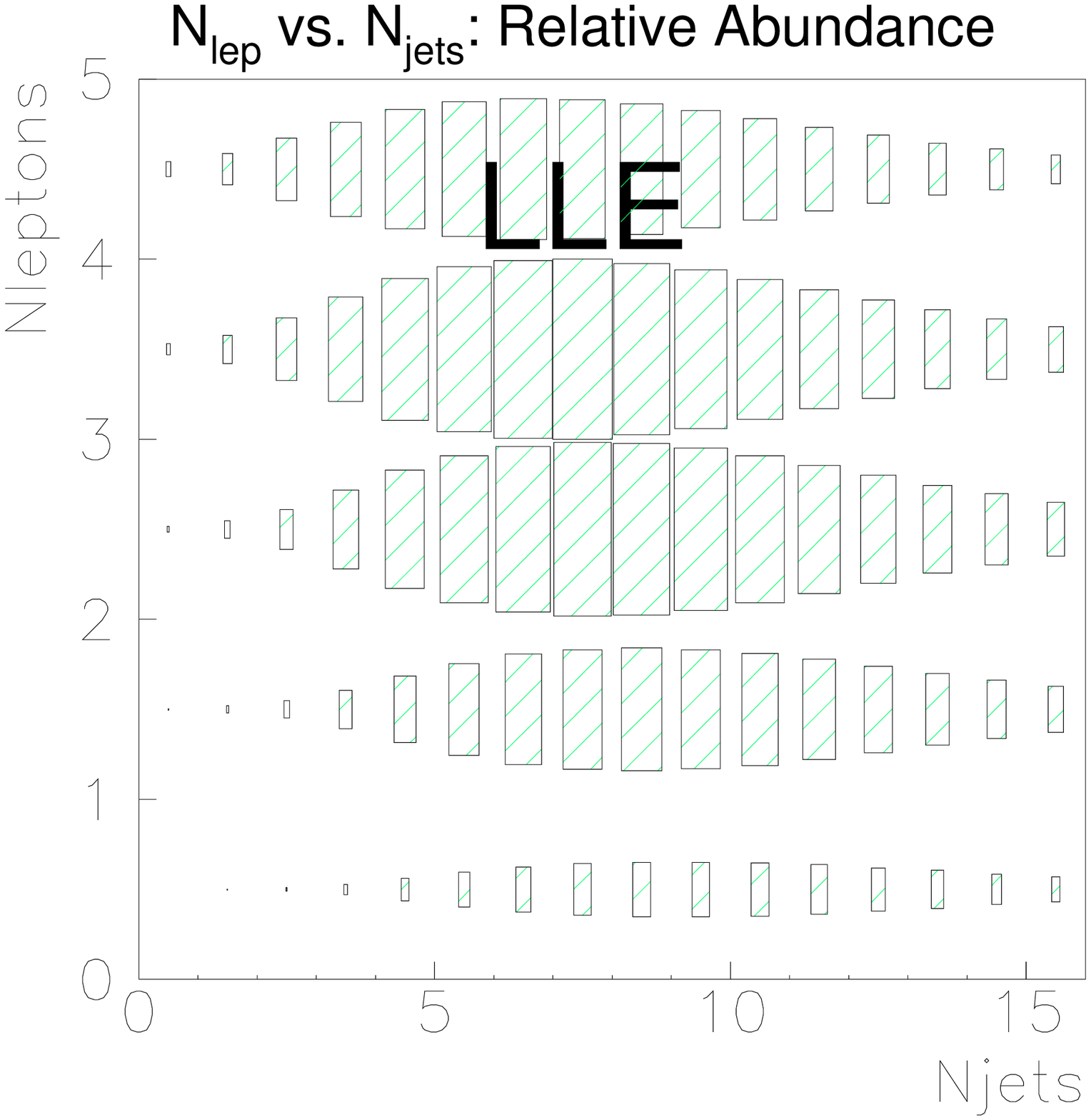}
\includegraphics*[scale=0.2]{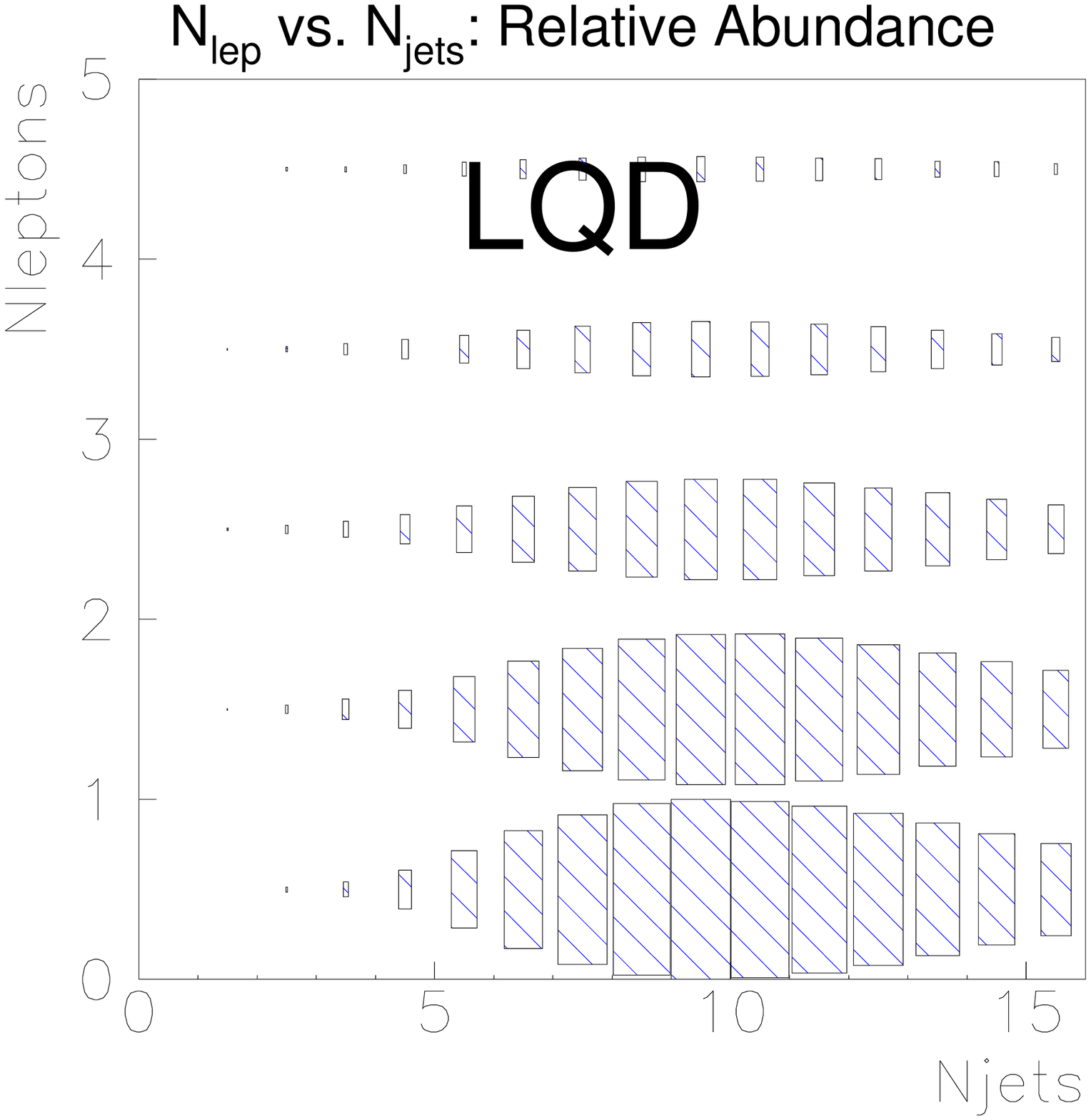}
\vspace*{-3mm}\end{center}
\caption{
Lepton versus jet multiplicity (see text) 
in mSUGRA $P_9$ 
for events surviving the cut on \ET. The numbering of the bins is such that
the events with 0 jets are in the
bin to the right of the number 0 and events with 0 leptons are in the bin
above the number 0. 
\label{fig:jl_p9}}
\end{figure}
\subsection{Leptons and Jets}
Due to the fact that the LSP decays, 
one expects an increased number of leptons and/or jets in the \LV-SUSY
scenarios. In figures \ref{fig:jl_sm} 
and \ref{fig:jl_p9} the number of leptons is plotted versus the
number of jets for each of the SM backgrounds and for $P_{9}$(MSSM), $P_{9a}$(LLE), and
 $P_{9a}$(LQD). All events satisfy $\ET>100\GeV$.

The absolute normalizations are, of course, very different for each of the SM
plots. 

One notices that the number of reconstructed jets goes up to 15 on the plots in
figs.~\ref{fig:jl_sm} and \ref{fig:jl_p9}. The jet numbers
are obtained using the \atlfast\ standard cone algorithm jet finder,
employing a cone size of $\Delta R=\sqrt{(\Delta \phi)^2+ (\Delta\eta)^2}=0.4$
and with clusters up to pseudorapidities $|\eta|<5$ included only if their
transverse energies are above $E_T^{min}=10\GeV$. Whether
a resolution of up to 15 jets or more 
is possible (and reliable) in the finished experiment is
questionable, yet the important point remains that there is more
hadronic activity associated with the LQD scenarios than is the case for the
MSSM, and we should be able to distinguish between the two possibilities
using any jet finder.

The increased number of leptons in the LLE scenario as well as the
increased number of jets (and a small increase in the number of leptons) in
the LQD scenario relative to the MSSM are evident. Since the $Z/W$ and
high-$p_T$ QCD backgrounds have the highest cross sections, a cut removing
events in the lower left corner was performed, requiring
$N_{\mathrm{Jets}}+N_{\ell}\ge 8$ or $N_\ell\ge 3$. This cut has a rejection factor of
about 20 for the SM events and efficiencies above 75\% for all \LV-SUSY
scenarios, again with a tradeoff between signal loss at low-mass points and
efficient background rejection for high-mass points. 
\subsection{Additional Variables}
In addition to the above mentioned requirements, 
cuts on several other kinematical variables
were performed before the neural network analysis below. We shall here only briefly
define and comment these variables (see \cite{skands_thesis} for details):
\begin{itemize}
\item The $p_T $ of the hardest object (jet or lepton) in the event was
  required to be greater than 200\GeV.
\item We define the ``$p_T$-weighted 4-object energy correlation'' by: 
\begin{equation}
P_{4C} \equiv \frac{1}{12}\left(\frac{E_4}{E_3}+\frac{E_3}{E_2} + \frac{E_2}{E_1}\right)(
p_{T1}+p_{T2}+p_{T3}+p_{T4})
\end{equation}
where $E_1,...,E_4$ ($p_{T1},...,p_{T4}$) are the energies (transverse
momenta) of the four hardest objects in the event. For pair-produced LSP's,
one would expect four hard objects with more or less equal energies more often than
would happen for background events. The energy fractions are weighted by the
average $p_T$ to give the variable some extra sensitivity against $t\bar{t}$
events. A cut at $P_{4C}>200\GeV$ was placed.
\item Events with Thrust greater than 0.85 were rejected. The Thrust
  calculation naturally suffers from the loss of particles down the beam-pipe
  at high pseudorapidities. Only particles with $|\eta|<5$ are defined as
  being inside the active calorimetry range in \atlfast, and only these
  particles are included in the Thrust calculation.
\item Events with Oblateness greater than 0.4 were
  rejected\footnote{Note that Oblateness is calculated entirely in the transverse
    plane for hadron colliders.}.
\item Events with Circularity less than 0.1 were rejected\footnote{Same
    comment as for Oblateness.}.
\end{itemize} 
\subsection{Neural Network Cuts}
It is clear that the above cuts do not define a dedicated search
strategy. They were applied to all scenarios with no distinction between
MSSM, LLE, and LQD models. A more comprehensive analysis would of course have
to focus on each of these possibilities separately, and also some
differentiation between the mSUGRA parameters would be required. In
particular sparticles lighter than the top and ones heavier than the
top could with advantage be searched for using separate strategies. 

This highlights the fact that the mSUGRA parameter space, including now the
\LV\ couplings, is not small. It was judged an uneconomical use of resources at this
point to optimize a purely physics-based analysis for each of the many
possibilities offered by this large parameter space. Instead, 
three neural networks were trained with post-trigger
events to separate MSSM, LLE, and LQD scenarios, respectively, from the
background distributions. Each network was thus trained not with one scenario
but with equal numbers of events from 15 different scenarios with varying
mSUGRA parameters and varying \LV-coupling strengths, 
allowing the networks to pick out general
qualities common to each class without over-fitting to a particular
model.

This procedure is certainly prone to the dangers which are always 
present when using neural networks on simulated data, in that the input
parameters (\emph{simulated} experimental observables) do not come with
warning labels about where in their domains the simulated results can be
trusted and where not. However, exercizing caution on this point, 
the method proposed here could be a useful alternative
for exploring theories with large parameter spaces.

As inputs were used all of 
the above mentioned variables along with the $p_T$ of the four hardest jets
and the two hardest leptons. 
The networks employed were single-layer perceptrons using the gradient
descent learning algorithm, with biased linear input and output neurons, and
hidden neurons activated according to a biased sigmoid (`logistic') function. At the end
of each learning period, the networks were trimmed using the OBD (Optimal
Brain Damage) prescription \cite{lecun90}. After ten such periods, the
networks generally showed a negligible difference in performance on the
learning sample as compared with the performance on an 
independent monitor sample, indicating
that over-fitting to the training samples is not a problem.

The training samples consisted of
3000 post-trigger SUSY events, selected equally among the appropriate
scenarios, and 4000 background events, weighted according to their
post-trigger cross sections to represent approximately $10^{7}$ events in the
learning algorithm. An output of zero (one) was the target for background
(SUSY) events. 

The finished LLE (LQD) network showed a clear preference for low-Thrust,
multi-lepton (multi-jet), low-Oblate-ness 
events with a mixture of low and high \ET. For
slightly higher Thrust values, low lepton momenta were preferred in the LLE
case. For the LQD network, the most noteworthy additional feature was a 
strong relaxation of all other considerations for very high jet
energies. Note that the background levels at low Thrust values suffer from
some theoretical uncertainty due to the fact that parton showers have been
used to generate the multi-jet final states, an approach which in the past 
has been known to give too few multi-jet events. Though  
it is not clear that present generators suffer from the same problems, 
it is important to note that there is a non-negligible and potentially dangerous
uncertainty in any analysis relying on parton-shower simulations for large jet
numbers and low Thrust values.

Processing now the full event samples remaining
after the cuts described above through each of the three (MSSM, LLE, and LQD)
networks, a final cut was made requring outputs larger than 0.9. Since the learning
samples were small compared to the full event samples (e.g.\ about 100.000 SUSY events
were generated per scenario and only a few hundred used for training) no
effort was made to exclude those events which had participated in the
learning process from the analysis. This is, in principle, a source of error,
yet we permit ourselves to 
ignore it due to the smallness of the learning samples and since
any problems related to over-fitting have been minimized by OBD.
For the SM backgrounds, the generated event samples had been
depleted considerably by the initial cuts, and only the $t\bar{t}$ and double
gauge boson samples had any 
events remaining at all after the network cuts (note that the learning
samples used for training -- for both background and SUSY -- 
were post-trigger events). 

The procedure used to construct
upper bounds on the actual event numbers was, for $N$ events remaining in a
sample, to calculate the mean, $\mu$, of the Poisson distribution which 
has exactly 5\% chance to result in $N$ or fewer events. $\mu$ is then
interpreted as a ``95\% CL'' upper bound on the number of events which could
have passed the cut. For the low-$p_T$ QCD events, this number was then subjected to
the same rejection factors under cuts as the high-$p_T$ sample (later the
rejections for the $t\bar{t}$ sample were used for both), and the
double gauge boson rejection factors were used for the upper bound on single gauge
events.

Typically, for 30\fb$^{-1}$, 
between 500 and 1000 signal events remain after cuts for $P_2$,
$P_{12}$, and $F_2$. Many more of course remain for $P_9$ because of the
larger cross section, but there is no hope for $P_7$ with only
114 events expected in total in 30fb$^{-1}$ of data. Yet one should bear in
mind that single sparticle production which has not been included here could
significantly increase the cross section if the \RV\ couplings are not much
smaller than the gauge couplings. For a hadron machine like the LHC, this
effect would only be big for the LQD terms since single slepton resonances
would then be possible. Of course, if \BV\ terms are present, single
squark production would be possible as well.

\subsection{Results}
We define the statistical significance with which a discovery can be made by
\begin{equation}
P = \frac{S}{\sqrt{S+B}} \label{eq:P}
\end{equation}
and conservatively 
interpret $P>5$ rather than the conventional $S/\sqrt{B}>5$ to mean that
a 5$\sigma$ discovery will be possible. Using the event numbers obtained in
the analysis (the estimates on $B$ being 95\% CL upper bounds) results in the
$P$ values listed in table \ref{tab:discoverypotential}. 

In reality, $P$ depends on (unknown) systematic QCD uncertainties (parton
distributions etc) and should be corrected for
the effects of pile-up, and so we can only be confident that a $5\sigma$
discovery is possible if $P$ is somewhat larger than 5. 
Therefore, aside from working with the definition, eq.~(\ref{eq:P}), 
we attempt to define a more pessimistic quantity in a very crude, \emph{ad
  hoc} manner which we shall call $P_{corr}$.

The non-inclusion of pile-up results in too optimistic estimates of $S/B$.
To include an estimate of the
reduction of this ratio, we rewrite eq.~(\ref{eq:P}) to:
\begin{equation}
P = \frac{\sqrt{S}}{\sqrt{1+B/S}}
\end{equation}
where we now include the effects of pile-up by multiplying $B/S$ by some
factor. That twice as many background events per signal event could be
passing the analysis if pile-up was included seems a reasonably pessimistic
guess. 

Furthermore, assuming that the intrinsic uncertainty on both $S$ and
$B$ coming from uncertainties on QCD parameters will, 
to a first approximation, work in the same direction and
with a comparable magnitude for
both $B$ and $S$, we expect that the denominator in the above formula is not
affected by this uncertainty, and so we 
include the QCD-related uncertainties by reducing the number of
signal events in the numerator by a factor of 1.5, believing this to be an
adequate worst-case estimate. This yields the following
form for the ``corrected discovery potential'':
\begin{equation}
P_{corr} = \frac{S}{\sqrt{1.5S + 3B}}\label{eq:Pcorr}
\end{equation}
Of course, this quantity should not be taken too seriously. We list it in 
table \ref{tab:discoverypotential} merely to show the effects of
the stated factors on the discovery potential, i.e.\ a reduction of $S/B$ by
a factor of 2 combined with a reduction of both $S$ and $B$ by 2/3.

With regard to the sensitivity of this analysis on systematic uncertainties
on the normalization of $B$, note that only for $P_9$ do we have an $S/B\gg
1$. For the $P_2$, $P_{12}$, and $F_2$ scenarios $S/B$ lies between one half
and unity. We do not estimate this to be a serious problem since our
background estimate is only a 95\% CL upper limit (larger event samples would
most likely bring the high-$p_T$ QCD component down) 
and since the background normalization can presumably be determined to better
than a factor 2 (as is used in $P_{corr}$) using complementary 
regions of phase space where the SUSY population is low or vanishing.

\begin{table*}
\begin{center}
{\hspace*{-0.4cm}
{\large\sf ATLAS \LV-SUSY DISCOVERY POTENTIAL}\\
{\hspace*{-0.4cm}\setlength{\extrarowheight}{3.pt}\sf
\begin{tabular}[t]{c|c}\toprule
\begin{tabular}[t]{lccc}
& \multicolumn{3}{c}{\sf NETWORK}\\
SUSY & MSSM & LLE & LQD \vspace*{-\extrarowheight}\\
Point &                         $P/P_{corr}$ & $P/P_{corr}$ & $P/P_{corr}$
\\ \cmidrule{1-4}
$ P_{2a}^{\mbox{\tiny LLE    }}$
&
$ 24.3
/
 16.2$
&
$ 25.5
/
 17.1$
&
$ 25.4
/
 17.1$
\\
$ P_{2b}^{\mbox{\tiny LLE    }}$
&
$ 24.5
/
 16.3$
&
$ 25.8
/
 17.3$
&
$ 25.7
/
 17.3$
\\
$ P_{2n}^{\mbox{\tiny LLE    }}$
&
$ 23.2
/
 15.4$
&
$ 24.7
/
 16.5$
&
$ 24.3
/
 16.3$
\\
$ P_{7a}^{\mbox{\tiny LLE    }}$
&
$  0.7
/
  0.4$
&
$  0.7
/
  0.4$
&
$  0.8
/
  0.5$
\\
$ P_{7b}^{\mbox{\tiny LLE    }}$
&
$  0.8
/
  0.4$
&
$  0.8
/
  0.4$
&
$  0.8
/
  0.5$
\\
$ P_{7n}^{\mbox{\tiny LLE    }}$
&
$  0.7
/
  0.4$
&
$  0.7
/
  0.4$
&
$  0.8
/
  0.5$
\\
$ P_{9a}^{\mbox{\tiny LLE    }}$
&
$191
/
153$
&
$315
/
256$
&
$218
/
176$
\\
$ P_{9b}^{\mbox{\tiny LLE    }}$
&
$190
/
153$
&
$316
/
256$
&
$218
/
176$
\\
$ P_{9n}^{\mbox{\tiny LLE    }}$
&
$166
/
133$
&
$257
/
208$
&
$169
/
135$
\\
$P_{12a}^{\mbox{\tiny LLE    }}$
&
$ 23.4
/
 15.5$
&
$ 25.6
/
 17.2$
&
$ 25.5
/
 17.2$
\\
$P_{12b}^{\mbox{\tiny LLE    }}$
&
$ 23.4
/
 15.5$
&
$ 25.5
/
 17.2$
&
$ 25.5
/
 17.2$
\\
$P_{12n}^{\mbox{\tiny LLE    }}$
&
$ 21.8
/
 14.4$
&
$ 24.2
/
 16.2$
&
$ 24.3
/
 16.3$
\\
$ F_{2a}^{\mbox{\tiny LLE    }}$
&
$ 11.3
/
  7.0$
&
$ 14.0
/
  8.8$
&
$ 13.3
/
  8.4$
\\
$ F_{2b}^{\mbox{\tiny LLE    }}$
&
$ 11.2
/
  6.9$
&
$ 13.7
/
  8.7$
&
$ 13.1
/
  8.2$
\\
$ F_{2n}^{\mbox{\tiny LLE    }}$
&
$  9.9
/
  6.1$
&
$ 12.4
/
  7.8$
&
$ 12.3
/
  7.7$
\\
\vspace*{-3ex}\\
\cmidrule{1-4}
$ P_{2a}^{\mbox{\tiny LQD    }}$
&
$ 20.9
/
 13.7$
&
$ 24.3
/
 16.2$
&
$ 24.8
/
 16.6$
\\
$ P_{2b}^{\mbox{\tiny LQD    }}$
&
$ 21.4
/
 14.1$
&
$ 24.7
/
 16.5$
&
$ 25.3
/
 17.0$
\\
$ P_{2n}^{\mbox{\tiny LQD    }}$
&
$ 21.5
/
 14.1$
&
$ 23.3
/
 15.5$
&
$ 24.2
/
 16.2$
\\
$ P_{7a}^{\mbox{\tiny LQD    }}$
&
$  1.0
/
  0.6$
&
$  1.0
/
  0.6$
&
$  1.1
/
  0.6$
\\
$ P_{7b}^{\mbox{\tiny LQD    }}$
&
$  1.0
/
  0.6$
&
$  1.0
/
  0.6$
&
$  1.1
/
  0.7$
\\
$ P_{7n}^{\mbox{\tiny LQD    }}$
&
$  1.0
/
  0.6$
&
$  1.0
/
  0.6$
&
$  1.1
/
  0.6$
\\
$ P_{9a}^{\mbox{\tiny LQD    }}$
&
$116
/
 91.6$
&
$153
/
122$
&
$125
/
 99.0$
\\
$ P_{9b}^{\mbox{\tiny LQD    }}$
&
$113
/
 88.7$
&
$151
/
121$
&
$123
/
 97.6$
\\
$ P_{9n}^{\mbox{\tiny LQD    }}$
&
$113
/
 88.5$
&
$131
/
104$
&
$110
/
 86.6$
\\
$P_{12a}^{\mbox{\tiny LQD    }}$
&
$ 15.7
/
 10.0$
&
$ 19.2
/
 12.5$
&
$ 20.9
/
 13.7$
\\
$P_{12b}^{\mbox{\tiny LQD    }}$
&
$ 15.8
/
 10.1$
&
$ 19.4
/
 12.6$
&
$ 21.1
/
 13.9$
\\
$P_{12n}^{\mbox{\tiny LQD    }}$
&
$ 16.6
/
 10.6$
&
$ 18.6
/
 12.1$
&
$ 20.9
/
 13.8$
\\
$ F_{2a}^{\mbox{\tiny LQD    }}$
&
$  7.0
/
  4.2$
&
$  9.5
/
  5.9$
&
$ 10.5
/
  6.5$
\\
$ F_{2b}^{\mbox{\tiny LQD    }}$
&
$  7.0
/
  4.2$
&
$  9.4
/
  5.8$
&
$ 10.5
/
  6.5$
\\
$ F_{2n}^{\mbox{\tiny LQD    }}$
&
$  6.9
/
  4.2$
&
$  8.8
/
  5.4$
&
$ 10.3
/
  6.4$
\\
\vspace*{-3ex}\\
\end{tabular}
&
\begin{tabular}[t]{lccc}
& \multicolumn{3}{c}{\sf NETWORK}\\
 SUSY & \multicolumn{1}{c}{MSSM} & 
\multicolumn{1}{c}{LLE} &
\multicolumn{1}{c}{LQD}\vspace*{-\extrarowheight}\\
Point                           
&$P/P_{corr}$ 
&$P/P_{corr}$ &
$P/P_{corr}$  
\\ \cmidrule{1-4}
$ P_{2a}^{\mbox{\tiny LLE+LQD}}$
&
$ 24.3
/
 16.2$
&
$ 25.9
/
 17.4$
&
$ 25.8
/
 17.4$
\\
$ P_{2b}^{\mbox{\tiny LLE+LQD}}$
&
$ 24.5
/
 16.3$
&
$ 25.9
/
 17.4$
&
$ 25.9
/
 17.4$
\\
$ P_{2n}^{\mbox{\tiny LLE+LQD}}$
&
$ 21.4
/
 14.0$
&
$ 23.2
/
 15.4$
&
$ 24.2
/
 16.2$
\\
$ P_{7a}^{\mbox{\tiny LLE+LQD}}$
&
$  0.8
/
  0.5$
&
$  0.8
/
  0.5$
&
$  0.9
/
  0.5$
\\
$ P_{7b}^{\mbox{\tiny LLE+LQD}}$
&
$  0.8
/
  0.5$
&
$  0.8
/
  0.5$
&
$  0.9
/
  0.5$
\\
$ P_{7n}^{\mbox{\tiny LLE+LQD}}$
&
$  1.0
/
  0.6$
&
$  1.0
/
  0.6$
&
$  1.1
/
  0.6$
\\
$ P_{9a}^{\mbox{\tiny LLE+LQD}}$
&
$179
/
143$
&
$289
/
234$
&
$203
/
164$
\\
$ P_{9b}^{\mbox{\tiny LLE+LQD}}$
&
$178
/
143$
&
$291
/
236$
&
$204
/
164$
\\
$ P_{9n}^{\mbox{\tiny LLE+LQD}}$
&
$114
/
 89.3$
&
$132
/
105$
&
$111
/
 87.0$
\\
$P_{12a}^{\mbox{\tiny LLE+LQD}}$
&
$ 20.7
/
 13.6$
&
$ 23.8
/
 15.8$
&
$ 24.3
/
 16.3$
\\
$P_{12b}^{\mbox{\tiny LLE+LQD}}$
&
$ 20.8
/
 13.6$
&
$ 23.8
/
 15.8$
&
$ 24.4
/
 16.4$
\\
$P_{12n}^{\mbox{\tiny LLE+LQD}}$
&
$ 16.5
/
 10.6$
&
$ 18.5
/
 12.0$
&
$ 20.8
/
 13.7$
\\
$ F_{2a}^{\mbox{\tiny LLE+LQD}}$
&
$  9.0
/
  5.5$
&
$ 11.9
/
  7.4$
&
$ 11.9
/
  7.5$
\\
$ F_{2b}^{\mbox{\tiny LLE+LQD}}$
&
$  9.0
/
  5.5$
&
$ 11.7
/
  7.3$
&
$ 11.8
/
  7.4$
\\
$ F_{2n}^{\mbox{\tiny LLE+LQD}}$
&
$  7.1
/
  4.3$
&
$  8.9
/
  5.5$
&
$ 10.4
/
  6.4$
\\
\vspace*{-3ex}
\\ \cmidrule{1-4}
$ P_{2 }^{\mbox{\tiny MSSM   }}$
&
$ 10.4
/
  6.4$
&
$ 10.2
/
  6.3$
&
$ 10.2
/
  6.3$
\\
$ P_{7 }^{\mbox{\tiny MSSM   }}$
&
$  0.2
/
  0.1$
&
$  0.2
/
  0.1$
&
$  0.2
/
  0.1$
\\
$ P_{9 }^{\mbox{\tiny MSSM   }}$
&
$136
/
108$
&
$121
/
 95.9$
&
$ 93.5
/
 72.8$
\\
$P_{12 }^{\mbox{\tiny MSSM   }}$
&
$ 16.1
/
 10.3$
&
$ 15.5
/
  9.9$
&
$ 16.1
/
 10.3$
\\
$ F_{2 }^{\mbox{\tiny MSSM   }}$
&
$  9.4
/
  5.8$
&
$  9.7
/
  6.0$
&
$ 10.5
/
  6.5$
\\
\vspace*{-3ex}
\end{tabular}\\
\bottomrule
\end{tabular}}}
\caption[\small ATLAS Discovery Potential]{ATLAS discovery potential and
corrected discovery potential (see text) for all
SUSY scenarios investigated using each of the three networks. 
The numbers shown correspond to an integrated luminosity of 30\fb$^{-1}$.
\label{tab:discoverypotential}}
\end{center}
\end{table*}

What one can see in table \ref{tab:discoverypotential} is again that the LLE
scenarios typically yield much purer event samples than the LQD ones, due
to the hadronic environment at the LHC. It is also worth noticing that it is
not easy, from this analysis alone, to discriminate between MSSM, LLE, and
LQD scenarios. This is due to the fact that the three networks were trained
independently, i.e.\ they were taught to reject SM events but not events from
the other scenario types. 

A suggestion for how we might have gained further insight
would be to construct \emph{one} network
with one output node for each class of models, 
requiring an output of (0,0,...) for background
events, (1,0,0,...) for events of class 1, (0,1,0,0,...) 
for events of class 2, and so on. One would
then be able to construct a measure for the relative probability of 
the events surviving cuts belonging to 
a given class of models, by comparing their $n$-dimensional distribution in
the network output space with the expected background
distribution. 

\section{Outlook and Conclusion \label{sec:conc}}
\subsection{Outlook:}
Though some preliminary studies have been performed in the present work, the
field is large and many important tasks remain. 
Recently, 
a significant  theoretical effort has been dedicated to
studying the case of baryon number violation
\cite{baer97,allanach01,allanach01-2}. 
Both single sparticle
production and baryon number violating SUSY will be included in the \pythia\
generator in the foreseeable future. 

This will hopefully spur more LHC activity, with 
comprehensive studies of both lepton and baryon number violating scenarios.
Preliminary studies indicate a lessening of the reach of the LHC in baryon
number violating scenarios \cite{baer97}, and it would be of interest to
explore this reach with the full range of production and decay mechanisms
included. In addition, studies should certainly be
carried out for mSUGRA benchmark points without neutralino LSP's.

Finally, triggers dedicated for \RV-SUSY should be incorporated in the
standard ATLAS trigger programme.
\subsection{Conclusion:}
1278 decay modes of Supersymmetric particles into Standard Model
particles through lepton number violating
couplings have been implemented in the \pythia\ event
generator. Combining this augmented version of the generator with a crude
simulation of the ATLAS detector, trigger menus for mid-luminosity running of
the LHC have been proposed and seen to have a high acceptance of supersymmetric
events in several $L$-violating SuperGravity scenarios while still giving
event rates in the 1Hz region. 

Taking these trigger menus as basis, the possibility for a 5$\sigma$
discovery after 30\fb$^{-1}$ data taking was estimated for each investigated
model, using a technique based on simultaneous classification of groups of
models rather than detailed studies of single models. 

For cross sections down to \tn{-10}\mb\ (or, roughly, for points
with $m_0<2\TeV$ and $m_{1/2}<1\TeV$) it was
found that a $5\sigma$ discovery was possible for all scenarios with
30\fb$^{-1}$ of data. It is estimated that uncertainties related to QCD
parameters or pile-up in the detector, both of which have not been taken into
account in the present analysis, could not significantly affect this conclusion. 

$L$-violating decays of the gluino and the possibility of
invoking $B$-violation will be implemented in the \pythia\ generator within a
foreseeable future.
\paragraph{Acknowledgement:} The author gratefully acknowledges support 
from Frederikke L\o rup f.\ Helms
Mindelegat and thanks P.\ Richardson and S.\ Mrenna for helpful correspondence.
\bibliography{main}
\end{document}